\documentclass[useAMS,usenatbib]{mnras}
\usepackage{ifthen,amsmath,graphicx,bm,amsbsy,aas_macros,url,array,amssymb,url,multirow,comment}\usepackage{aas_macros,todonotes}
\include{amsmath,graphicx}
\begin{document}

\title[Constraining High Redshift X-ray Sources]{Constraining High Redshift X-ray Sources with Next Generation 21\,cm Power Spectrum Measurements}

\author[Ewall-Wice et al.]{Aaron Ewall-Wice$^{1}$ \thanks{E-mail: aaronew@mit.edu},
Jacqueline Hewitt$^{1}$,
Andrei Mesinger$^{2}$,
Joshua S. Dillon$^{1,3}$,
\newauthor
Adrian Liu$^{4,\dagger}$,
Jonathan Pober$^{5}$,
\\
$^{1}$Dept. of Physics and MIT Kavli Institute, Massachusetts Institute of Technology, Cambridge, MA 02139, USA \\
$^{2}$Scuola Normale Superiore, Piazza dei Cavalieri 7, 56126 Pisa, Italy\\
$^{3}$Dept. of Astronomy, UC Berkeley, Berkeley CA 94720\\
$^{4}$Berkeley Center for Cosmological Physics, University of California, Berkeley, Berkeley, CA 94720\\
$^{\dagger}$Hubble Fellow. \\
$^{5}$Brown University, Department of Physics, Providence, RI 02912, USA}

\maketitle

\begin{abstract}
	We use the Fisher matrix formalism and semi-numerical simulations to derive quantitative predictions of the constraints that power spectrum measurements on next-generation interferometers, such as the Hydrogen Epoch of Reionization Array (HERA) and the Square Kilometre Array (SKA), will place on the characteristics of the X-ray sources that heated the high redshift intergalactic medium. Incorporating observations between $z=5$ and $z=25$, we find that the proposed 331 element HERA and SKA phase 1 will be capable of placing $\lesssim 10\%$ constraints on the spectral properties of these first X-ray sources, even if one is unable to perform measurements within the foreground contaminated ``wedge" or the FM band. When accounting for the enhancement in power spectrum amplitude from spin temperature fluctuations, we find that the observable signatures of reionization extend well beyond the peak in the power spectrum usually associated with it. We also find that lower redshift degeneracies between the signatures of heating and reionization physics lead to errors on reionization parameters that are significantly greater than previously predicted. Observations over the heating epoch are able to break these degeneracies and improve our constraints considerably. For these two reasons, 21\,cm observations during the heating epoch significantly enhance our understanding of reionization as well. 
\end{abstract}

\voffset-.6in


\section{Introduction}

Observations of 21\,cm emission from neutral hydrogen at high redshift will provide us with a unique window into the  cosmic dark ages, the birth of the first stars, and the reionization of the IGM (see \citet{Furlanetto:2006Review,Morales:2010,Pritchard:2012} for reviews). Current experiments aiming to detect the cosmological 21\,cm line currently follow two tracks. The first involves accessing the sky averaged ``global signal" with a single dipole which is being attempted by experiments such as EDGES \citep{Bowman:2010}, LEDA \citep{GreenHill:2012,Bernardi:2015}, DARE \citep{Burns:2012}, SciHi \citep{Voytek:2014}, and BIGHORNS \citep{Sokolowski:2015}. Alternatively, one can observe spatial fluctuations in emission with a radio interferometer  and  new generation of instruments are already collecting data with the aim of making a first statistical detection. These include the MWA \citep{Tingay:2013a}, PAPER \citep{Parsons:2010}, LOFAR \citep{VanHaarlem:2013}, and the GMRT \citep{Pagica:2013}. Already, a number of upper limits have been established over the redshifts at which reionization is expected to have occurred  \citep{Dillon:2014,Parsons:2014,Jacobs:2015,Ali:2015,Dillon:2015b}. This first generation of instruments may possess the sensitivity to make a low signal to noise detection of the power spectrum during the Epoch of Reionization (EoR). However, planned experiments such as the Square Kilometer Array (SKA) and the Hydrogen Epoch of Reionization Array (HERA) will be capable of constraining the parameters in reionization models to several percent precision \citep{Pober:2014,Greig:2015a}. 

The 21\,cm line is an extremely rich observable, extending far beyond the start of reionization into the dark ages. The EoR is only the final chapter in the remarkable story that it encodes. Before reionization, X-rays emitted from the first generations of high mass X-ray binaries (HMXB) \citep{Mirabel:2011} and/or the hot interstellar medium (ISM) \citep{Pacucci:2014} reversed the adiabatic cooling of the IGM, likely bringing it into emission against the cosmic microwave background (CMB). Experiments will observe the brightness temperature of 21\,cm emission against the CMB which is given by \citep{Furlanetto:2006Review}
\begin{equation}
 \delta T_b \approx 9.2 x_{HI}( 1 + \delta) \left[ 1 - \frac{T_{cmb}}{T_s}\right] \left[ \frac{H(z)/(1+z)^{1/2}}{dv_\|/dr} \right] \text{mK},
\end{equation}
where $x_{HI}$ is the neutral fraction, $\delta$ is the fractional baryon overdensity, $T_\text{cmb}$ is the temperature of the CMB at the redshift of emission ($z$), $T_s$ is the spin temperature of the HI gas, $H(z)$ is the hubble parameter at redshift $z$, and $dv_\|/dr$ is the radial velocity gradient. 
At the end of adiabatic cooling, the $(1-T_\text{cmb}/T_s)$ factor is relatively large and negative \citep{Furlanetto:2006Global}. The first sources heating the IGM with X-rays introduce large dynamic range in this factor that is significantly greater in absolute separation than the permitted $[0,1]$ in $x_{HI}$. This leads to a power spectrum that is roughly an order of magnitude larger than during reionization except in extreme heating models \citep{Pritchard:2007,Santos:2008,Baek:2010,Mesinger:2013}. 

Recent work has shown that inefficient heating scenarios might be constrained with lower redshift measurements on current experiments \citep{Christian:2013,Mesinger:2014}. Indeed, the latest PAPER upper limits have ruled out an IGM with a spin temperature below $\approx 10$\,K \citep{Pober:2015a,Greig:2015b}. While X-rays have some effect on the ionization field \citep{Mesinger:2013}, their primary impact is encoded in  $T_s$ in all but the most extreme cases. Assuming that the linear relation between X-ray luminosity and star formation rate is similar to what is observed locally. $T_s$ is expected to be much greater than $T_\text{cmb}$, during reionization, eliminating its impact on $\delta T_b$ \citep{Furlanetto:2006Global}. Hence, to precisely determine the physics of IGM heating, measurements at redshifts higher than reionization may be needed. What limits on heating are possible at low redshift and to what extent measurements of the pre-reionization epochs are necessary to establish precision limits are open questions. 

In this paper, we use semi-numerical simulations and the Fisher matrix formalism to determine the limits that next generation experiments will put on the properties of X-ray sources during the cosmic dawn. In particular, we focus on what we might expect to learn from observations of the reionization epoch alone and what measurements at higher redshifts  might add to our knowledge. We also explore what additional reionization constraints might exist at pre-reionization redshifts. Earlier studies of constraints on reionization parameters \citep{Pober:2014,Greig:2015a} assume that $T_s \gg T_\text{cmb}$ which \citet{Mesinger:2013} have shown can significantly under-predict the power spectrum amplitude early on and before reionization. Back-lit by a negative $(1-T_\text{cmb}/T_s)$, there may exist additional detectable reionization signatures at higher redshift.

This paper is organized as follows. In \S~\ref{sec:Fisher} we describe the Fisher matrix formalism which we use to connect uncertainties in power spectrum measurements error bars on IGM heating properties. In \S~\ref{sec:MethodsSims} we describe our semi-numerical simulations along with our model parameters and their fiducial values. We describe our calculations of thermal noise in \S~\ref{sec:MethodsNoise} and discuss the arrays examined. In \S~\ref{sec:Derivatives} we examine the information encoded in the derivative of the power spectrum with respect to each astrophysical parameter. We discuss the degeneracies between parameters and our predictions for model limits in \S~\ref{sec:Constraints} and conclude in \S~\ref{sec:Conclusions}.

\section{From Measurements to Constraints}\label{sec:Fisher}
Today, the quantity that interferometers are attempting to estimate is $\Delta^2(k)$, the power spectrum of 21\,cm brightness temperature fluctuations. The power spectrum at comoving wavenumber $k=|{\bf k}|$ is defined in terms of the spatial Fourier transform of the brightness temperature field, $\widetilde{\delta}_b({\bf k})$.
\begin{equation}
\left \langle \widetilde{\delta}_b({\bf k}) \widetilde{\delta}_b({\bf k'}) \right \rangle = \delta_D({\bf k} - {\bf k'})\frac{2 \pi^2}{k^3} \Delta^2(k),
\end{equation}
where $\langle \rangle$ denotes the ensemble average over all realizations of the brightness field and $\delta_D$ is the Dirac delta function.

21\,cm observations of $\Delta^2$ can be connected to theory and simulations by performing a maximum likelihood (ML) analysis to obtain the best-fit parameters for an astrophysical model. The uncertainty on the parameters obtained from such an analysis may be approximated using the Fisher matrix formalism which we now describe.

The Fisher matrix describes the amount of information, contained within a data set, on the parameters of a model. It can be defined through the derivatives of the ln-likelihood about the ML parameter values \citep{Fisher:1935}.
\begin{equation}\label{eq:Fisher}
F_{ij} \equiv -\left \langle \frac{\partial^2 \ln \mathcal{L} }{\partial \theta_i \partial \theta_j}  \right \rangle.
\end{equation}
Here, $\mathcal{L}$ is the likelihood of observing the outcome of a measurement given a set of model parameters ${\bf \theta}$. In our case the measurement is the set of power spectrum values in each Fourier and redshift bin while ${\bf \theta}$ is the set of parameters in our astrophysical model. Intuitively, we see in equation~\ref{eq:Fisher} that the largest amount of information exists for parameters that cause the greatest change in the likelihood.

If the likelihood function is Gaussian, which is usually a good approximation in the case of small errors about the ML point, then the covariance matrix of these parameters , $C_{ij}$, is simply the inverse of the Fisher matrix, 
\begin{equation}
C_{ij} = \left({\bf F^{-1}}\right)_{ij}. 
\end{equation}
The more strongly $\mathcal{L}$ depends on $\theta_i$ and $\theta_j$, the smaller the covariance matrix values and the error bars on our estimates. 

 Judicious choices in constructing the estimate of the power spectrum from visibility data can ensure that the Fourier modes and redshift bins at which the power spectrum is estimated are uncorrelated \citep{Liu:2011,Dillon:2013}. Assuming that these errors are Gaussian, the ln-likelihood is given by
 \begin{equation}
 \ln \mathcal{L}({\bf x},\boldsymbol{\theta}) = - \sum_\beta \frac{1}{2 \sigma_\beta^2} ( x_\beta - \Delta^2_\beta (\boldsymbol{ \theta}) )^2,
 \end{equation} 
 where $\beta$ indexes each (k,z) bin and $\sigma_\beta$ is the standard deviation of the power spectrum in the $\beta^{th}$ bin. Taking the derivatives with respect to $\theta_i$ and $\theta_j$, one obtains
\begin{equation}\label{eq:FisherSum}
F_{ij} = \sum_\beta \frac{1}{\sigma_\beta^2} \frac{\partial \Delta^2_\beta}{\partial \theta_i} \frac{\partial \Delta^2_\beta}{\partial \theta_j} + \frac{2}{\sigma_\beta^2}\frac{\partial \sigma_\beta}{\partial \theta_i} \frac{\partial \sigma_\beta}{\partial \theta_j}.
\end{equation}
 The first term in this formula is identical to the equation appearing in \citet{Pober:2014} and arises from the mean term in the power spectrum estimate.  The second term involves the derivative of the standard deviation of each power spectrum bin and accounts for the fact that cosmic variance is proportional to the power spectrum itself. 
\footnote{Formulas similar to equation~\ref{eq:FisherSum} have been derived in the context of CMB analysis \citep{Bunn:1995,Vogeley:1996,Tegmark:1997d}. In the CMB case, the observable is usually taken to be the complex $a_{\ell m}$ which are distributed as a zero-mean Gaussian. In this case, the information in the variance of these terms leads to a formula very similar to the first sum in equation~\ref{eq:FisherSum} (e.g. equation~18 in \citet{Tegmark:1997d}) while the means contribute nothing. }

To elucidate the relative contribution from the second term in equation~\ref{eq:FisherSum} to $F_{ij}$, we may assume that our measurement of $\Delta_\beta^2$ is dominated by sample variance which is a decent approximation for the modes next generations arrays such as HERA-331 and the SKA will be most sensitive to \citep{Mesinger:2014}. One can choose bin-sizes small enough such that $\Delta^2$ is effectively constant over each bin, hence $\sigma_\beta = \Delta_\beta^2/\sqrt{N_\beta}$ where $N_\beta$ is the number independent measurements in the $\beta^{th}$ bin. With this approximation, equation~\ref{eq:FisherSum} becomes
\begin{equation}
F_{ij} = \sum_\beta \frac{N_\beta}{\Delta_\beta^4} \left(1 + \frac{2}{N_\beta} \right) \frac{\partial \Delta_\beta^2}{\partial \theta_i} \frac{\partial \Delta_\beta^2}{\partial \theta_j}.
\end{equation}
Hence the contribution to the Fisher information from sample variance is only significant for small $N_\beta$. For the arrays studied in this paper, we find that the variance term in equation~\ref{eq:FisherSum} tends to only contribute at $1\%$ the level of the mean term so we ignore it and obtain an identical expression to the one appearing in \citet{Pober:2014}. Dropping the variance information allows us to write equation~\ref{eq:FisherSum} as the dot product between two vectors used in \citet{Pober:2014}
\begin{equation}
F_{ij} \equiv {\bf w}_i \cdot {\bf w}_j,
\end{equation}
where
\begin{equation}
w_i(k,z) = \frac{1}{\sigma(k,z)}\frac{\partial \Delta^2(k,z)}{\partial \theta_i}.
\end{equation}
Inspecting $w_i$ allows us to determine the contribution of each (k,z) bin to the Fisher information on each parameter.


\section{Simulations of the 21\,cm Signal}\label{sec:MethodsSims}
The formalism that we set out in \S~\ref{sec:Fisher} requires two ingredients: simulations of the brightness temperature field as a function of redshift and calculations of the error bars on each power spectrum measurement. In this section we address the first ingredient. We first briefly describe the publicly available 21cmFAST\footnote{\url{http://homepage.sns.it/mesinger/DexM___21cmFAST.html}} code which we use to simulate the signal (\S~\ref{ssec:21cmFAST}). We then describe the specific reionization and heating parameters that we choose to vary and our choices for their fiducial values (\S~\ref{ssec:Params}).
\subsection{Semi-Numerical Simulations}\label{ssec:21cmFAST}

We generate realizations of $\delta T_b$ using the {\tt 21cmFAST} code for which a detailed description is available in \citet{Mesinger:2007} and \citet{Mesinger:2011}. The simulation volume is 750\,Mpc on a side with $400^3$ cells. Here we give a brief overview of its treatment of the heating and reionization physics.

 A density field is calculated at each redshift by evolving an initial Gaussian random field via the Zeldovich approximation. An ionization field is then computed from the density fluctuations using the excursion set formalism of \citet{Furlanetto:2004}. Cells of the density field, filtered at a comoving scale $R$, are determined to be ionized if the fraction of mass that has collapsed into virialized structures, $f_{coll}$, exceeds the inverse of an ionizing efficiency parameter, defined as $\zeta$. $R$ is varied between the pixel size up to the mean free path (MFP) of ultraviolet (UV) photons in the HII regions which we denote as $R_\text{mfp}$.
 
The HI spin temperature couples to both the kinetic temperature of the gas (through collisions) and the Lyman-$\alpha$ flux from the first generations of stars via the Wouthuysen-Field effect \citep{Field:1959}. In most scenarios, Lyman-$\alpha$ coupling saturates early on while the high optical depth to Lyman-$\alpha$ absorption couples the color temperature of UV photons to the kinetic temperature of the HI gas. Hence, $T_s$ primarily reflects the thermal state of the HI gas and astrophysical processes affecting it.  The impact on the kinetic temperature from X-rays is determined by integrating the X-ray specific emissivity along a light cone for each cell. The specific emissivity is assumed to be dominated by HMXBs or hot ISM and hence proportional to the star formation rate \citep{Mineo:2012a,Mineo:2012b}. The full expression for the emissivity in each simulation cell used in 21cmFAST is \citep{Mesinger:2013}
\begin{align}\label{eq:Emissivity}
\epsilon_{h_p\nu}(\nu ,{\bf x}, z') & = \alpha_X h_p \frac{f_X}{10} \left( \frac{\nu}{\nu_\text{min}} \right)^{-\alpha_X} \rho_\text{SFR}({\bf x},z')  
\end{align}
For $\nu \ge \nu_\text{min}$ and $0$ otherwise. Here, $h_p$ is the planck constant, $\nu_\text{min}$ is a low-energy obscuration threshold, $\alpha_X$ is the spectral index of X-ray emission and $f_X$ is known as the ``X-ray efficiency" which serves as an overall normalization parameter with $f_X \approx 1$ corresponding to $0.1$ X-ray photon per baryon involved in star formation. Note that the factor of $\nu_\text{min}$ within the exponent removes its effect on the overall X-ray luminosity. The star formation rate density, $\rho_{SFR} ( {\bf x},z')$, is approximated within each voxel with the equation \citep{Mesinger:2013}
\begin{equation}
\rho_{SFR}({\bf x},z') = \left[\langle \rho_b \rangle f_* (1+ \delta_{nl}({\bf x} )) \frac{df_{coll}}{dt}({\bf x}) \right]
\end{equation}
where $\langle \rho_b \rangle$ is the mean baryon density, $f_*$ is the fraction of baryons in stars (assumed to be $0.1$), $\delta_{nl}$ is the non-linear overdensity averaged over all smoothing scales, and $d f_{coll}/dt$ the derivative of the fraction of mass collapsed into viarialzed objects with virial temperatures greater than a certain threshold (which we explain below), It is computed using the hybrid prescription of \citet{Barkana:2004}. 

\subsection{Astrophysical Parameters and their Fiducial Values}\label{ssec:Params}
In this work, we explore our ability to constrain a six-parameter model of reionization and heating which accounts for the major astrophysical degrees of freedom, $(\zeta,R_\text{mfp},T_\text{vir}^\text{min},f_X,\nu_\text{min},\alpha_X)$ which are defined below. We choose a dimensionless parameterization by letting $\theta_i$ be the fractional difference of the parameter from its fiducial values. For example, $\theta_\zeta \equiv (\zeta-\zeta^\text{fid})/ \zeta^\text{fid}$. We label the set of $\theta$s for $(\zeta,R_\text{mfp},T_\text{vir}^\text{min},f_X, \nu_\text{min}, \alpha_X)$ as $(\theta_\zeta,\theta_R,\theta_T,\theta_f,\theta_\nu,\theta_\alpha)$.

 The sensitivity of 21\,cm experiments to the first three of these parameters was previously considered in \citet{Pober:2014} and \citet{Greig:2015a}. These works only considering reionization redshifts and ignored the contribution to $\delta T_b$ from spin temperature fluctuations. \citet{Mesinger:2013} show that $f_X$ and other heating parameters have an impact during the early stages of reionization, potentially introducing previously ignored degeneracies. 21\,cm measurements during the heating epoch may therefor enhance our understanding of reionization. We now describe our choices for the fiducial value at which we compute the derivatives in equation~\ref{eq:FisherSum} for each parameter.

\begin{itemize}

\item  $\boldsymbol{\zeta} {\bf:}$ The ionization efficiency describes the number of ultraviolet  photons per unit time that enter the IGM from galaxies. It primarily affects the timing of reionization and can be written as a degenerate combination of parameters, $\zeta = f_{esc}f_* N_{\gamma/b} (1+n_{rec})^{-1}$ \citep{Furlanetto:2004}, where $f_{esc}$ is the UV escape fraction, $N_{\gamma/b}$ is the number of ionizing photons per baryon in stars and $n_{rec}$ is the number of times a typical HI atom undergoes recombination. Combinations of the optical depth to the CMB, quasar absoption data, the kinetic Sunyaev-Zeldovich effect allow some constraints on $\zeta$ \citep{Mesinger:2012} and is typically thought to lie between $5$ and $50$. We choose a fiducial value of $\zeta=20$ which, combined with our other fiducial choices, predicts a mean optical depth to the CMB of $\tau_e \approx 0.08$, and 50\% reionization at $z_{re}=8.5$ which is in line with current constraints from \citet{Ade:2015}.

\item ${\bf R_\text{mfp}} {\bf :}$ The MFP of UV photons in HII regions, this parameter determines the maximum HII bubble size and primarily determines the location of the ``knee" in the power spectrum. Physically, $R_\text{mfp}$ is set by the number density and optical depth of Lyman-limit systems. $R_\text{mfp}$ is highly unconstrained with limits from observations of Lyman-$\alpha$ systems at $z\sim 6$ allowing for values between $3$ and $80$\,Mpc \citep{Songaila:2010}. Subgrid modeling of inhomegenous recombination point towards a smaller $R_\text{mfp}$ between $5$ and $20$\,Mpc \citep{Sobacchi:2014} leading us to choose a fiducial value of $R_\text{mfp}=15$\,Mpc.

\item ${\bf T_{vir}^{min}} {\bf :}$ The emission lines of H$_2$ are expected to serve as the predominant cooling pathway for primordial gas, allowing for the collapse and fragmentation necessary for star formation. While radiative cooling is possible in halos with $T_\text{vir}\gtrsim 100$K \citep{Haiman:1996a,Tegmark:1997c}, photodissociation induced by UV background set up by the first sources suppresses cooling in less massive halos without sufficient self shielding \citep{Haiman:1996b,Haiman:1997}. Even if sufficient cooling occurs, feedback can also suppress star formation \citep{Springel:2003,Mesinger:2008,Okamoto:2008}. Studies indicate a broad range of plausible minimum virial temperatures between $T_\text{vir} = 10^2-10^5$\,K   corresponding to halo masses between $10^6-10^8$\,M$_\odot$ at $z\approx 10$. We adopt a fiducial $T_\text{vir}^\text{min}$ of $1.5 \times 10^4$\,K which corresponds to the threshold for atomic line cooling. $T_\text{vir}^\text{min}$ directly enters our simulations by determining the minimum mass above which $f_{coll}$ and $df_{coll}/dt$ is calculated. While thermal or mechanical feedback has the potential to change $T_\text{vir}^\text{min}$ as a function of redshift \citep{Mesinger:2013}, we assume that all the halos involved in heating the IGM also take part in reionizing it and hold $T_\text{vir}^\text{min}$ constant. We will sometimes group $T_\text{vir}^\text{min}$ with other ``reionization" parameter due to its inclusion in the three-parameter model of previous works that only address reionization \citep{Mesinger:2012,Pober:2014,Greig:2015a,Greig:2015b}. However, we emphasize that it determines the minimal masses of the halos driving X-ray heating as well and is just as much an ``X-ray heating parameter" as any of the parameters determining the X-ray SEDs of early galaxies which we now list.

\item ${\bf f_X :}$ Our fiducial value of $f_X=1$ is chosen to give an integrated 0.5-8\,keV luminosity of $\approx 5 \times 10^{39}$ erg s$^{-1}$ M$_{\odot}$ yr$^{-1}$, consistent with the $\approx 3 \times 10^{39}$ erg s$^{-1}$ M$_{\odot}$ yr$^{-1}$ observed at z=0 by \citet{Mineo:2012a} and corresponds to 0.1 X-ray photons per stellar baryon. While $f_X=1$ matches local observations, there are reasons to expect different values at high redshift. For example, the decrease in metallicity with redshift might increase the rate of very X-ray luminous black hole X-ray binaries, boosting $f_X$ \citep{Mirabel:2011}.  Observations out to $z\sim 4$ on {\it Chandra Deep Field-South} \citep{Xue:2011} have been interpreted with conflicting results.  \citet{Cowie:2012} do not find evolution in the X-ray luminosity to star formation rate out to redshift $\sim 4$ while \citet{Basu-Zych:2013} claim to observe weak evolution consistent with population synthesis models \citep{Fragos:2013} after adjusting for dust extinction. 

\item $\boldsymbol{\nu}_{\bf min} {\bf: }$ Absorption by the ISM in early galaxies will cause the emergent spectral energy distribution to differ from the ones intrinsic to the sources. In particular, large $\nu_\text{min}$ can lead to a particularly hard X-ray spectrum that delays heating and reduces the contrast between hot and cold patches during heating \citep{Fialkov:2014}. The degree to which absorption is expected to be present depends critically on both the column density and composition of the ISM of the host galaxies with metals absorbing X-rays with energies above $0.5$\,keV and helium primarily absorbing softer X-rays. Our fiducial choice of the X-ray obscuration threshold, $\nu_\text{min}=0.3$~keV is identical to that used in \citet{Pacucci:2014} and describes an ISM with a similar column density to the Milky Way but with low metallicity.

\item $\boldsymbol{\alpha}_{\bf X} {\bf :}$ The spectral index of X-ray emission from early galaxies. X-ray emission from local galaxies is observed to originate from two different sources: X-ray binaries (XRB) and the diffuse hot ISM. XRB spectra generally follow a power law of $\alpha \sim 1.0$ between $0.5$ and $2$ keV \citep{Mineo:2012a}. Emission from the hot ISM originates from metal line cooling and thermal bremsstrahlung in the plasma generated by supernovae and stellar winds. The diffuse emission observed by \citet{Mineo:2012b} has been found by \citet{Pacucci:2014} to be well approximated by a power law of $\alpha \sim 3.0$ between $\sim 0.5-10$~keV. These authors also observed that the maximum amplitude of the power spectrum during X-ray heating is highly sensitive to $\alpha_X$. A steeper spectrum resulting from ISM dominated emission is abundant in soft X-rays and leads to significantly higher contrast between hot and cold regions due to their short MFP. This leads to a boost in power spectrum amplitude by a factor of $\sim 3$. We choose a fiducial $\alpha_X$ of $1.2$.

\end{itemize}

\begin{figure}\label{fig:Global}
\includegraphics[width=\columnwidth]{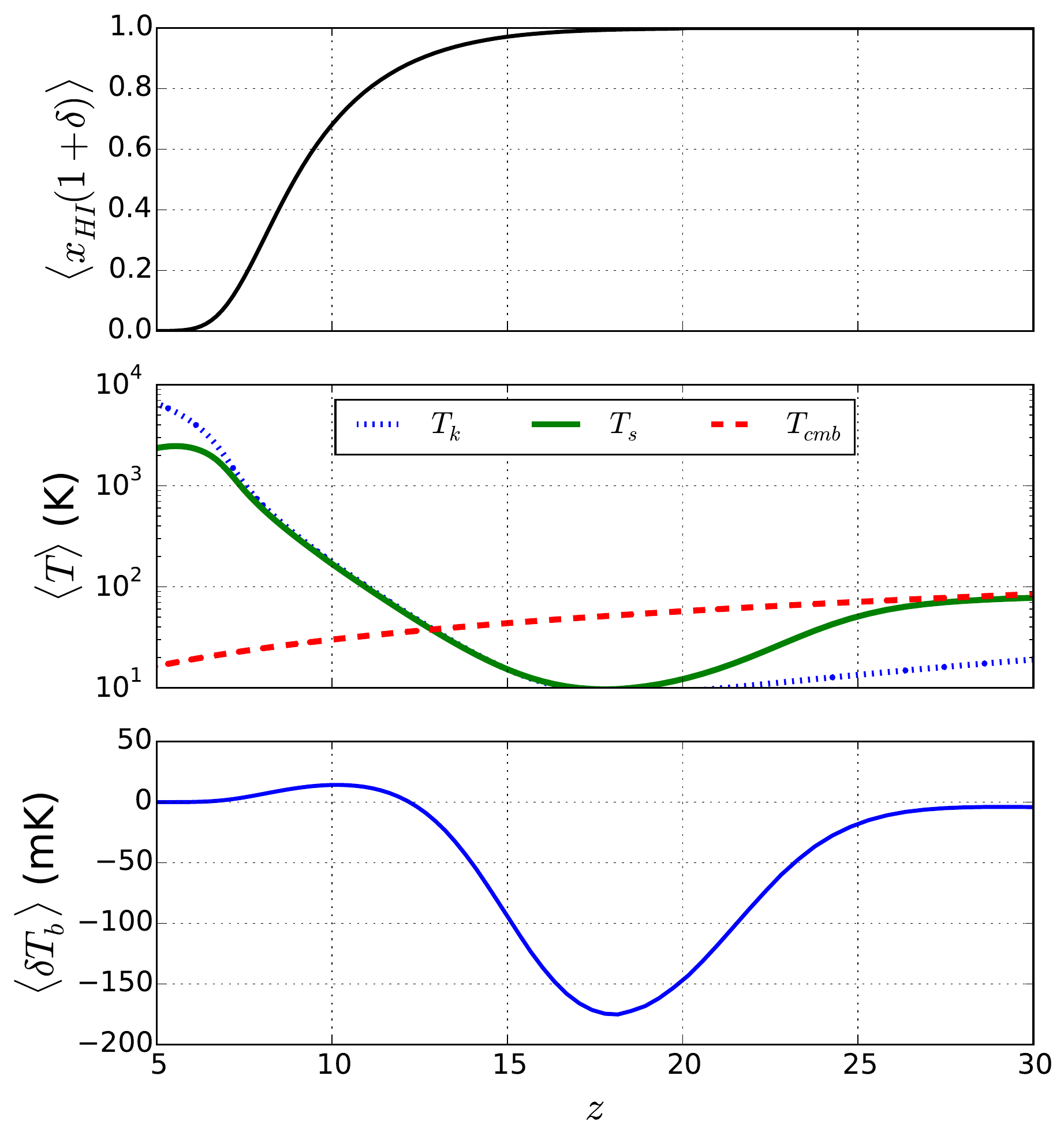}
\caption{Top: The evolution of the density weighted average of the neutral fraction. Middle: The evolution of the kinetic temperature ($T_k$), $T_s$, $T_\text{cmb}$. Bottom: the redshift evolution of the mean 21\,cm brightness temperature, $\langle \delta T_b \rangle $. All averages are taken over volume.}
\label{fig:Global}
\end{figure}

\begin{figure*}
\includegraphics[width=\textwidth]{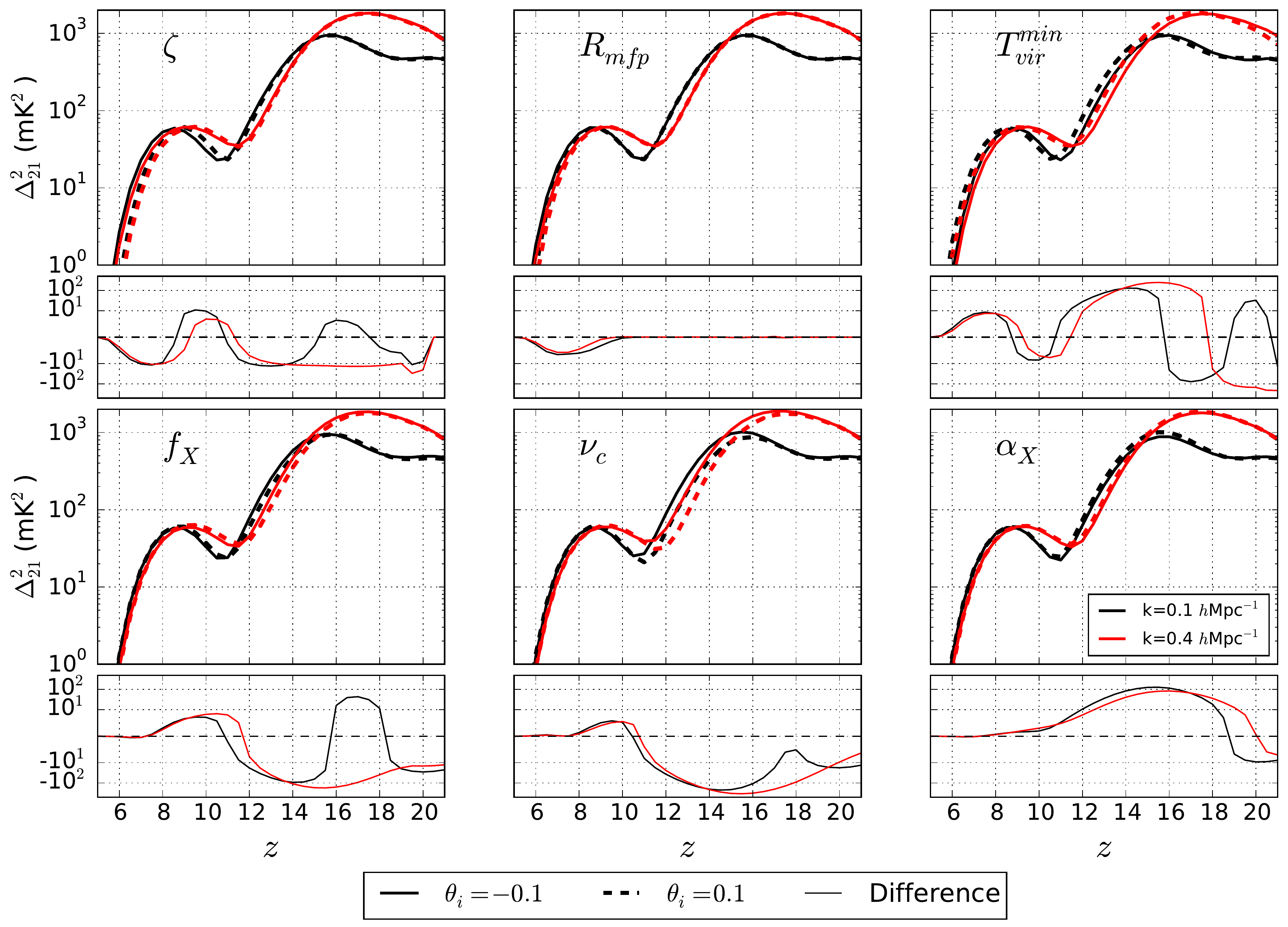}
\caption{$\Delta^2_{21}$ as a function of redshift for two different 1d k bins, $k=0.1h$Mpc$^{-1}$ (black lines) and $k=0.4h$Mpc$^{-1}$ (red lines). For each parameter, we show the power spectrum at $\theta = \pm 0.1$ (thick solid and thick dashed lines respectively) along with the difference (thin solid lines). Note that our parameterization defines $\theta$ as the fractional difference of each parameter from its fiducial. The first two peaks of the three-peaked structure, discussed in \citet{Pritchard:2007,Santos:2008,Baek:2010} and \citet{Mesinger:2013}, is clearly visible representing the epochs of reionization and X-ray heating. With the exception of $R_\text{mfp}$, parameter changes affect a broad range of redshifts.}
\label{fig:psParams}
\end{figure*}

For each astrophysical parameter, we run six simulations varying $\theta_i$ by $\pm 1\%$, $\pm 5\%$, and $\pm 10\%$ of its fiducial value and a linear fit of $\Delta^2(k,z)$ is used to compute $\partial \Delta^2 / \partial \theta_i$. Inspection shows that the power spectrum and is well described by a linear trend over this range of parameter values. In Fig.~\ref{fig:Global}, we show the evolution of the density-weighted-average of the neutral fraction as a function of redshift for our fiducial model which predicts $50\%$ reionization at $z_{re}\approx 8.5$. We also display the volume averaged evolution of the kinetic, spin, and 21\,cm brightness temperatures compared to the evolution of the CMB temperature. We see that $\langle T_s \rangle$ exceeds $T_\text{cmb}$ at $z \approx 12.5$  but is within an order of magnitude of $T_\text{cmb}$ out to a redshift of $z \approx 9$, hence we can expect some signatures of spin temperature fluctuations to be present in our signal early on in reionization. In Fig.~\ref{fig:psParams} we show our power spectra as a function of redshift at two different comoving scales at $\pm 10\%$ of their fiducial parameter values. The evolution of the power spectrum on large scales follows the three peaked structure noted in \citet{Pritchard:2007,Santos:2008,Baek:2010,Mesinger:2013} corresponding, in order of redshift, to reionization, X-ray heating, and Lyman-$\alpha$ coupling. The redshifts of these peaks depends on the comoving scale. At $k=0.1$, the peaks occur at $z=8.5,15,$and $22.5$.

\section{Instrument and Foreground Models}\label{sec:MethodsNoise}
The next ingredient for our Fisher matrix analysis is a set of error bars on each of the observed cosmological modes. The error on each spherically averaged power spectrum estimate not only depends on both instrumental parameters such as the collecting area, $uv$-coverage, and observing time, but also strongly on our ability to mitigate foreground contamination.

 We investigate two large experiments: the Hydrogen Epoch of Reionization Array (HERA), which is being commissioned in South Africa, and the Square Kilometer Array (SKA). For each instrument, we derive the power spectrum sensitivity using the public {\tt 21cmSense} code\footnote{\url{https://github.com/jpober/21cmSense}} \citep{Pober:2013b,Pober:2014} assuming 6 hours of observations per night over 180 days and a spectral resolution of $100$\,kHz. For simplicity, we also assume that all arrays perform drift-scan observations. While the SKA is not a drift-scan instrument, the sensitivity difference between a drift scan and tracking a single eight degree field for six hours per night was only found to be on the order of $\sim 10-20\%$ \citep{Pober:2015b} depending on the degree of foreground contamination. We assume that observations are taken simultaneously at all redshifts between 5 and 30 and that power spectra are estimated from sub-bands with a co-evolution bandwidth of $\Delta z = 0.5$, to avoid strong evolution effects\citep{Mao:2008}. The redshift interval assumes observing over the frequency band between $57$ and $237$\,MHz and we note that simultaneous observations over such a large band may not be possible with a single feed, raising the observing time by a factor of two. For each array, we consider a discrete $uv$ plane upon which measurements are gridded with a cell size set by the instrument's antennae footprint and using rotation synthesis, we compute the number of seconds of of observing performed in each cell $\tau({\bf k})$. We emphasize that $\tau({\bf k})$ is different for each observed frequency due to the fact that each instrumental baseline has a fixed physical length and antenna size while the $uv$ cell a baseline occupies is set by the number of wavelengths between its two antennas. The $uv$ cell size also depends on frequency since the number of wavelengths spanned by the physical antenna changes. The variance of the power spectrum estimate within each $uv$ cell is given by 
\begin{equation}\label{eq:noise}
\sigma^2({\bf k}) =  \left[ X^2 Y \frac{\Omega' T^2_{sys}}{2\tau({\bf k})}  + P_{21}({\bf k}) \right]^2,
\end{equation}
where $\Omega'$ is the ratio between the solid angle integral of the primary beam squared \citep{Parsons:2014} and the solid angle integral of the beam while $T_{sys}$ is the sum of the sky and receiver temperatures whose values we choose to be $T_{rec} = 100$\,K and $T_{sky} =60 \lambda^{2.55}$\,K \citep{Fixsen:2011}. 
$X$ is the comoving angular diameter distance and $Y$ is a linear conversion factor between frequency and radial distance given by \citep{Morales:2004}
\begin{equation}
Y=\frac{c(1+z)^2}{H_0 f_{21} E(z)},
\end{equation}
where $f_{21}\approx 1420$\,MHz is the frequency of 21\,cm radiation, $c$ is the speed of light, $H_0$ is the hubble parameter at $z=0$, and $E(z)=H(z)/H_0$. At each observed frequency and array, $\tau({\bf k})$ is computed by dividing the uv plane into cells with diameter $D/\lambda$ and adding up the cumulative time within each cell occupied by each array's antennas after rotation synthesis. $\Omega'$ is computed assuming a gaussian beam with $\sigma = 0.45 \lambda/D$ where $D$ is the diameter of the antenna element, an approximation that ensures the volume of the central lobe of the airy disk for the aperture matches that of a Gaussian. The sensitivity within each $k$-bin is computed by taking the inverse variance weighted average of all $uv$ cells within the bin that are not contaminated by foregrounds.

 While foregrounds are expected to dominate the signal by a factor of $\gtrsim 10^5$, they are also spectrally smooth and only occupy a limited region of k-space known as the ``wedge" \citep{Datta:2010,Parsons:2012,Morales:2012,Vedantham:2012,Thyagarajan:2013,Trott:2012,Hazelton:2013,Liu:2014a,Liu:2014b,Thyagarajan:2015a,Thyagarajan:2015b}. The degree to which we might be able to observe inside (and close to) this foreground contaminated region will depend crucially on our ability to characterize the foregrounds and our instrument. Since the extent of the wedge corresponds to the angular offset of sources from the phase center, our ability to characterize and subtract sources from the primary beam sidelobes will determine what uncontaminated modes will be available for power spectrum estimation as seen in \citet{Thyagarajan:2015a,Thyagarajan:2015b} and \citetext{Pober et al. Submitted}. We consider the three different foreground scenarios from \citet{Pober:2014} to describe the efficacy of foreground isolation. 
\begin{itemize}
\item {\bf Optimistic (Foreground Subtraction):} All modes outside the full width half power of the primary beam are sufficiently decontaminated as to be used in power spectrum estimation. We also include a small buffer of $k_\parallel=0.05$\,$h$Mpc$^{-1}$ to account for intrinsic spectral structure in the foregrounds and/or the instrument. This buffer is significantly smaller than the supra horizon emission observed in \citet{Pober:2013a} out to $\approx 0.1$\,$h$Mpc$^{-1}$. Ionospheric diffraction, whose severity runs inversely with frequency will likely increase the difficulty of foreground subtraction at X-ray heating redshifts.

\item {\bf Moderate (Foreground Avoidance):} In this scenario, we assume that all modes within the wedge are unusable and supra-horizon emission extends to $0.1$\,$h$Mpc$^{-1}$ beyond the wedge, in line with observations \citep{Pober:2013a,Parsons:2014,Ali:2015,Dillon:2015b}. 

\item {\bf Pessimistic (Instantly Redundant Delay Transform Power Spectrum):} This scenario is almost identical to our moderate foregrounds scenario except that only baselines that are instantaneously redundant in local sidereal time are added coherently. This is the sensitivity achievable using the current delay power spectrum technique \citep{Parsons:2012} which thus far has lead to the most stringent upper limits in the field \citep{Ali:2015}. We note that there is no fundamental reason for the delay transform technique to not coherently add partially redundant visibilities which is an ongoing topic of research. 
\end{itemize}
Both $\Omega'$ and $\tau({\bf k})$ in equation~\ref{eq:noise} depend on our instrument. In this work, we consider the following experimental configurations:
\begin{itemize}
\item {\bf HERA-127/331:} The Hydrogen Epoch of Reionization Array (HERA) is an experiment undergoing commissioning now, to detect the 21\,cm brightness temperature fluctuations during and before the EoR. Focusing on the foreground avoidance approach that has thus far proved successful for PAPER, HERA is designed to maximize collecting area outside of the wedge by filling the $uv$ plane with short baselines.  The antenna layout for HERA involves 331 hexagonally-packed, 14\,m diameter dishes. A staged buildout is expected to occur with the penultimate and ultimate stages comprising of a 127 and 331 dish core. HERA will also contain an additional 21 outrigger antennas to assist in imaging and foreground characterization. However, we do not include these outriggers in our analysis since they do not contribute significantly to HERA's sensitivity which is derived primarily from its short, core baselines. 
\item {\bf SKA-1 LOW:} We base our model of the SKA-LOW instrument on the description in \citet{Dewdney:2013}. We also reduce the antenna count by 50\% to reflect the recent rebaselining, making it nearly identical to the proposed design \#1 in \citet{Greig:2015c}. The array is comprised of 446, 35\,m diameter phased arrays of log-periodic dipole antennas. These stations are distributed in radius as a Gaussian with 75\% of antennae falling within a 1\,km radius. 
\end{itemize}

\section{Power Spectrum Derivatives and Their Physical Origin}\label{sec:Derivatives}

Having described our simulations of the signal and instrumental noise, we are in a position to discuss the two stages of our results which include the derivatives of the power spectrum with respect to each parameter and the resulting covariances. In this section we provide physical intuition for the outputs of our derivative calculations and the nature of the  information on each quantity that is available at different redshifts. 

\begin{figure*}
\includegraphics[width=.9\textwidth]{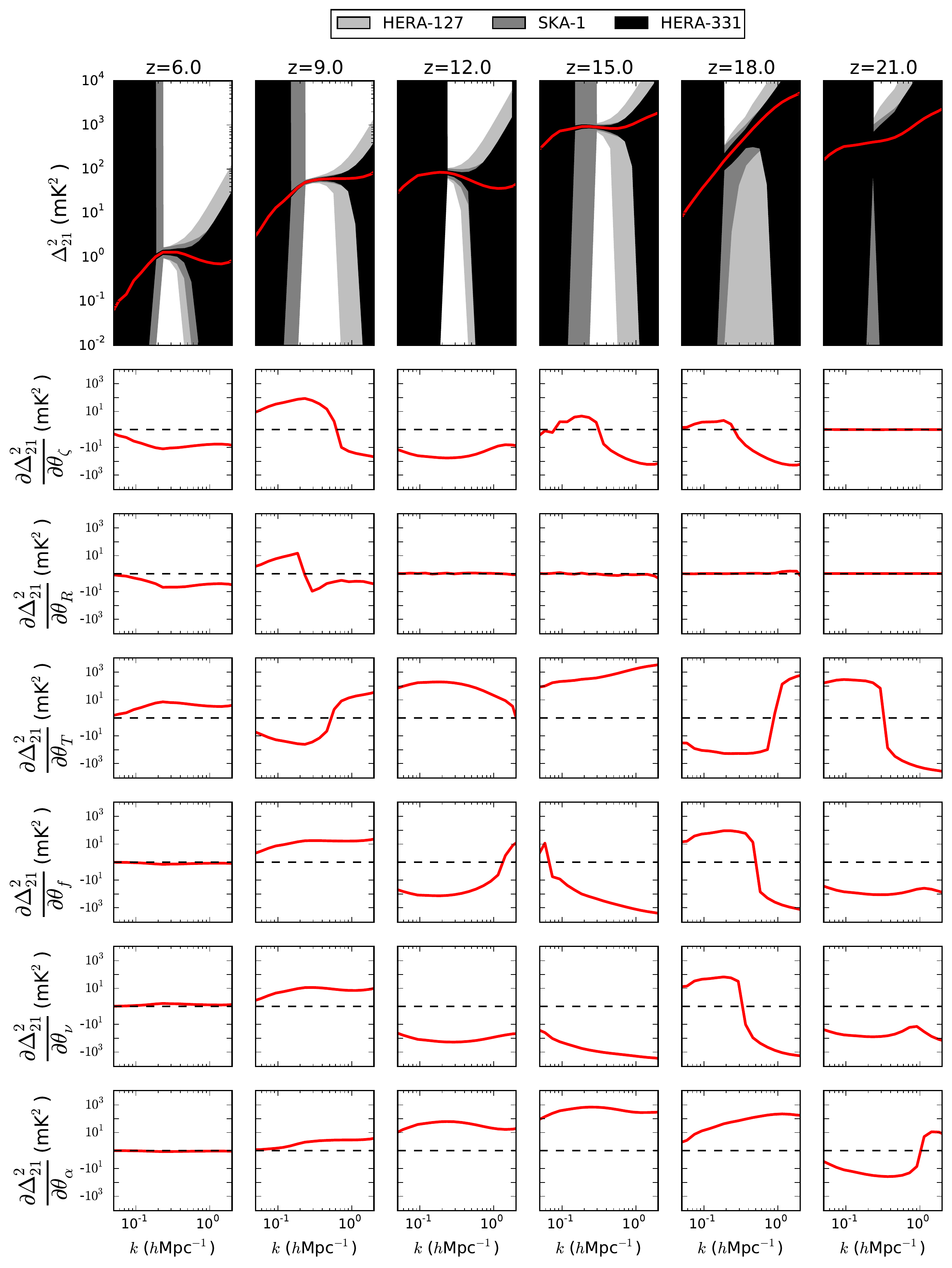}
\caption{Top: The power spectrum of 21\,cm fluctuations (solid red lines) over numerous redshifts. Filled regions denote the $1\sigma$ errors for the instruments considered in this paper with moderate foregrounds. Bottom: The derivatives of the 21\,cm power spectrum with respect to the astrophysical parameters considered in this work as a function of $k$ at various redshifts. Derivatives are substantial over all redshifts except for $R_\text{mfp}$ which only affects the end of reionization. Notably, $\partial_\zeta \Delta^2_{21}$ is negative on small scales at high redshift, a signature of the beginnings of inside out reionization while X-ray spectral parameters follow very similar redshift trends, indicative of degeneracy.}
\label{fig:Derivatives}
\end{figure*}

We show our fiducial power spectrum along with the $1\sigma$ uncertainty regions for the arrays studied in this paper at the top of Fig.~\ref{fig:Derivatives}. As observed in numerous previous works (e.g. \citet{Pritchard:2007,Christian:2013,Mesinger:2014}), the 21\,cm signal is detectable out to $z \approx 21$ due to the larger contrast available between cold and hot regions of the IGM during heating \citep{Mesinger:2013}. We show $\partial \Delta^2_{21}/\partial \theta_i$ for our astrophysical parameters at the bottom of Fig.~\ref{fig:Derivatives} and see that that with the exception of $R_\text{mfp}$, the derivative of the power spectrum with respect to the ionization parameters is substantial out to redshifts beyond the typical range associated with reionization. While we compute our Fisher matrix from redshift bins of $\Delta z=0.5$, for legibility, the panels in Fig.~\ref{fig:Derivatives} are shown for intervals of $\Delta z=3.0$. In addition, In the next two sections, we discuss, in detail, the origins of the trends in the derivatives related to both reionization (\S~\ref{ssec:infoIon}) and X-ray heating (\S~\ref{ssec:infoHeat}). 

\subsection{How Reionization Parameters Affect the 21\,cm Power Spectrum.}\label{ssec:infoIon}
We now describe the trends in the derivatives associated with reionization, several of which have already been discussed in the literature \citep{Mesinger:2012,Pober:2014,Sobacchi:2014} as well as new signatures present at high redshift that are only detectable with the inclusion of the large negative $(1-T_\text{cmb}/T_s)$ term supplied by the spin temperature calculation.

\begin{figure*}
\includegraphics[width=\textwidth]{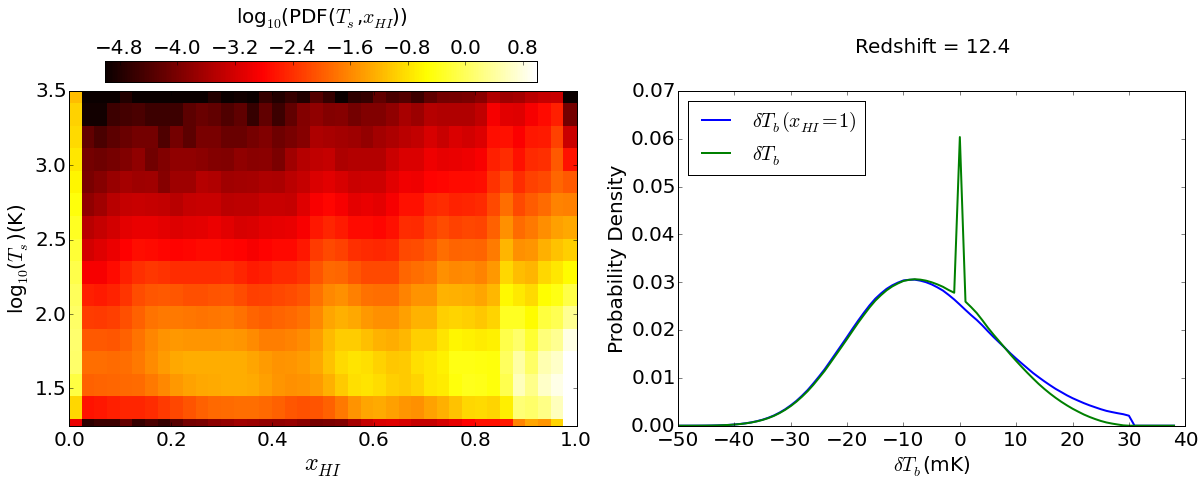}
\caption{Left: The logarithm of the probability distribution function (PDF) of pixels at $z=12.4$ for our fiducial model as a function of $T_s$ vs. $x_{HI}$. Even before the majority of reionization, early HII bubbles ionize the hottest points in the IGM, leading to a pileup of high $T_s$ pixels at $x_{HI}=0$ and reducing the contrast in $\delta T_b$ between hot and cold regions. Right: PDFs of $\delta T_b$ with and without $x_{HI}$ manually set to unity everywhere. The presence of ionization during X-ray heating leads to a decrease in the large $T_s$ wing, near $30$\,mK, and a spike at $0$\,mK, leading to a reduction in the dynamic range of the field and an overall decrease in power.}
\label{fig:xHContrast}
\end{figure*}

The primary effect of increasing $\zeta$ is to accelerate the time of reionization, shifting the peak to higher redshift. Thus we see a negative derivative to the left and the positive derivative to the right of the reionization peak. It is interesting to note that the derivative with respect to $\zeta$ remains significant (and primarily negative) far beyond the rise of the reionization peak and well into the heating epoch. This is due to the fact that the small precursor ionization bubbles exist out to high redshift, occupying the same voxels with the largest over-densities and greatest $T_s$. These HII bubbles set $\delta T_b$ to $0$ at the hottest points in the IGM and reduce the contrast between hot spots and the cold background (Fig.~\ref{fig:xHContrast}). Because these bubbles occur on small spatial scales, this leads to a reduction in the power spectrum amplitude with increasing $\zeta$ at large $k$. At large scales, we see a positive derivative at the rise of the heating peak and a negative derivative at its fall as we might expect if the heating rate were increased (Fig.~\ref{fig:psParams}). An increased ionization fraction decreases the optical depth of X-rays from the photoionization of HI, HeI, and HeII, providing such a rate increase.  The falling edge of the Lyman-$\alpha$ peak is similarly affected, perhaps due to an increase in the number of Lyman-$\alpha$ photons arising from X-ray excitations. 
	
Increasing the MFP of ionizing photons is known to  shift the ``knee" of the power spectrum to larger comoving scales. The diameter of the regions corresponding to our fiducial $R_\text{mfp}$ of $15$\,Mpc$^{-1}$ correspond to $k\approx 0.3$\,$h$Mpc$^{-1}$, hence the positive derivative below $k \approx 0.3$\,$h$Mpc$^{-1}$ and the negative derivative at smaller scales.  Since the MFP in HII regions only affects the brightness temperature fluctuations once the HII bubbles themselves have had time to grow out to this scale, we see no significant effect of $R_\text{mfp}$ on the early reionization power spectrum beyond $z \approx 9$. \citet{Mesinger:2012} note that a smaller $R_\text{mfp}$ has the effect of delaying the end of reionization, explaining the negative derivative across all scales at the lowest redshifts.

$T_\text{vir}^\text{min}$ affects both reionization and heating, however we will discuss its affect in this section. Since increasing $T_\text{vir}^\text{min}$ delays heating and reionization, its clearest signature is to shift the peaks towards low redshift, leading to positive differences to the left of each peak and negative differences to the right (Fig.~\ref{fig:psParams}). Smaller comoving scales ($k \gtrsim 0.4$\,$h$Mpc$^{-1}$) transition through the peaks at earlier times than larger scales (see Fig.~\ref{fig:psParams}). At $z=9$ and $18$, small comoving scales are at the fall of the reionization and heating peaks respectively (hence the positive derivatives) and at $z=21$ at the rise of the heating peak (rather than the fall of Lyman-$\alpha$) leading to the negative derivative at large $k$ in Fig.~\ref{fig:Derivatives}.

\subsection{How X-ray Spectral Properties Affect the Power Spectrum.}\label{ssec:infoHeat}

We now describe and provide physical intuition for the derivatives with respect to the X-ray spectral properties of galaxies before reionization.

As increasing $f_X$ raises the heating rate, the most obvious consequence is to shift the X-ray heating peak to higher redshift \citep{Mesinger:2013,Christian:2013,Mesinger:2014,Pacucci:2014}. In  Fig.~\ref{fig:Derivatives} this trend is clearly observable in the positive derivative at the rising edge of the heating peak and the negative derivative at the falling edge. There is also a significant positive derivative early in reionization and a slightly negative one at its conclusion. The reionization peak is impacted by a number of competing effects related to $f_X$. Since X-rays also generate secondary ionizations and have a longer MFP than UV photons, highly emissive scenarios produce a partially ionized haze that reduces the contrast between ionized and neutral patches and the amplitude of the reionization power spectrum \citep{Mesinger:2013}. Secondary ionizations also have the effect of shifting the reionization peak to higher redshift. As the power spectrum maximum occurs $z\approx z_{re}$, this causes an increase in power at the start of reionization and a decrease at the tail end. Finally, increasing $f_X$ raises the spin temperature during reionization. Since we are in the regime where $T_s > T_\text{cmb}$ over the reionization peak, increasing $f_X$ leads to an increase in the $(1-T_\text{cmb}/T_s)$ factor in $\delta T_b$, leading to an overall increase in the reionization power spectrum before the spin temperature's impact has saturated. In Figs~\ref{fig:psParams} and \ref{fig:Derivatives} we see the difference is positive during the onset of reionization and negative during the fall; indicating that the rise in spin temperature from increased $f_X$ and the shift to higher redshifts dominates the onset of reionization but that the direct spin temperature effects are saturated by the end.

We next examine the derivatives with respect to the obscuration threshold, $\nu_\text{min}$. Raising $\nu_\text{min}$ hardens the X-ray spectrum of the first galaxies. Since hard X-rays have significantly longer MFPs, this delays their absorption, leading to relatively uniform late heating \citep{Fialkov:2014,Pacucci:2014}. This has the effect of shifting the minimum between the reionization and heating peaks when $\langle \delta T_b \rangle \approx 0$ to lower redshift while suppressing the amplitude of heating fluctuations. In addition to the longer MFP, the heating delay is also due to harder X-rays depositing a larger fraction energy into ionizations rather than heating \citep{Furlanetto:2010}. The increased ionization energy fraction also leads to a slight reduction in the mean neutral fraction across redshift. One might also expect a slight shift in the reionization peak to high redshift as well and a decrease in amplitude from the reduced spin temperature and lower contrast in the ionization field from increased X-ray ionizations. While a shift and amplitude reduction in the heating peak is clear, we do not see a decrease in the reionization peak. In fact, inspection of Figs~\ref{fig:psParams} and ~\ref{fig:Derivatives} indicates that the amplitude of the peak actually increases. An explanation for this behavior is that the ionization and $(1-T_\text{cmb}/T_s)$ fields anti-correlate, leading to a negative contribution to the overall power spectrum amplitude. A larger $\nu_\text{min}$ leads to a decrease in the contrast between hot and cold patches which reduces the overall amplitude of this negative cross correlation and causes an increase in the reionization power spectrum. While the reduction in contrast raises the power spectrum amplitude in the neighborhood of our fiducial model, the trends towards a smaller reionization power spectrum dominate at much larger $\nu_\text{min}$ \citep{Mesinger:2013,Fialkov:2014}. 

As noted in \citet{Pacucci:2014}, increasing $\alpha_X$ reduces the mean free path of the X-rays, amplifying the constrast between hot and cool patches and leading to an increase in the heating power spectrum amplitude over the entire peak.  The increased spin temperature also drives up the amplitude of the ionization power spectrum and causes a slight shift towards higher redshifts as well. $\alpha_X$ has the weakest signature during the reionization epoch, perhaps in part due to the amplification of the anticorrelation between spin temperature and ionization fraction canceling out the increase in $T_s$. We note that in models with much lower heating efficiency that an increased $\alpha_X$ leads to a noticable dip during the rise of the reionization power spectrum (e.g. Fig.~8 in \citet{Christian:2013} or Fig.~5 in \citet{Pacucci:2014}) due to the enhancement of the $T_s$-$x_{HI}$ anti-correlation.

\section{Constraints from Heating Epoch Observations}\label{sec:Constraints}
	
	We now turn to the results of our Fisher matrix calculation. We focus our discussion on degeneracies and the dependence of overall constraints on the range of observed redshifts. 
	
\subsection{Degeneracies Between Parameters}\label{ssec:Degeneracy}

Combining our power spectrum derivatives with our sensitivity estimates, we calculate $w_i(k,z)$ which indicates the relative contribution of each 1d k bin and redshift to the Fisher information on each parameter. In addition, we can easily spot the sources of covariance between the different parameters by looking for similar $k$ and $z$ evolution. If at the same redshifts and $k$ values, two parameters have opposite (equal) signs in $w_i$, then a positive change in the first parameter can be compensated for by a positive (negative) change in the other, leading to degeneracy. This degeneracy can be broken if additional redshifts and Fourier modes are added in which the parameters do not have similar evolution. We denote the $w_i$s for $(\zeta,R_\text{mfp},T_\text{vir}^\text{min},f_X,\nu_\text{min},\alpha_X)$ as $(w_\zeta, w_R, w_T, w_f,w_\nu,w_\alpha)$. 
\begin{figure*}
\includegraphics[width=\textwidth]{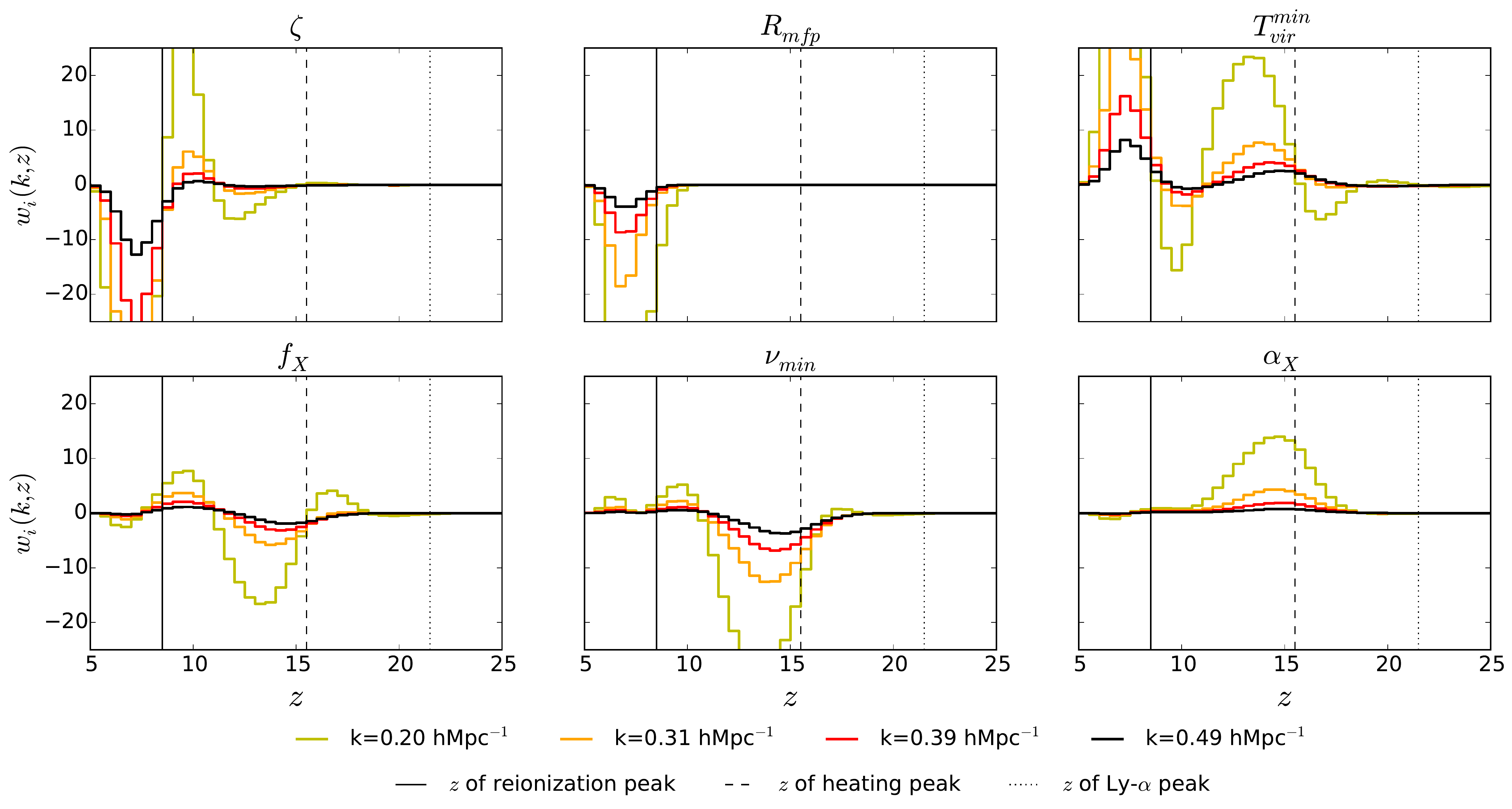}
\caption{Plotting $w_i(k,z)$ versus redshift for several different co-moving scales gives us a sense of the covariances between various parameters. Here we assume 1080 hours of observation on HERA-331 and the moderate foreground model. Since the thermal noise on $\Delta^2$ increases rapidly with $k$, $w_i$ is maximized at larger spatial scales. As we might expect from our fiducial model, $w_i$ for reionization parameters is maximized at lower redshifts while $w_i$ for X-ray spectral parameters is significant over the heating epoch. $T_\text{vir}^\text{min}$ affects both heating and reionization and has a broad redshift distribution. Vertical lines indicate the location of each of the three power spectrum peaks at $k=0.1h$\,Mpc$^{-1}$} 
\label{fig:wCompare}
\end{figure*}

\begin{figure}
\includegraphics[width=.48\textwidth]{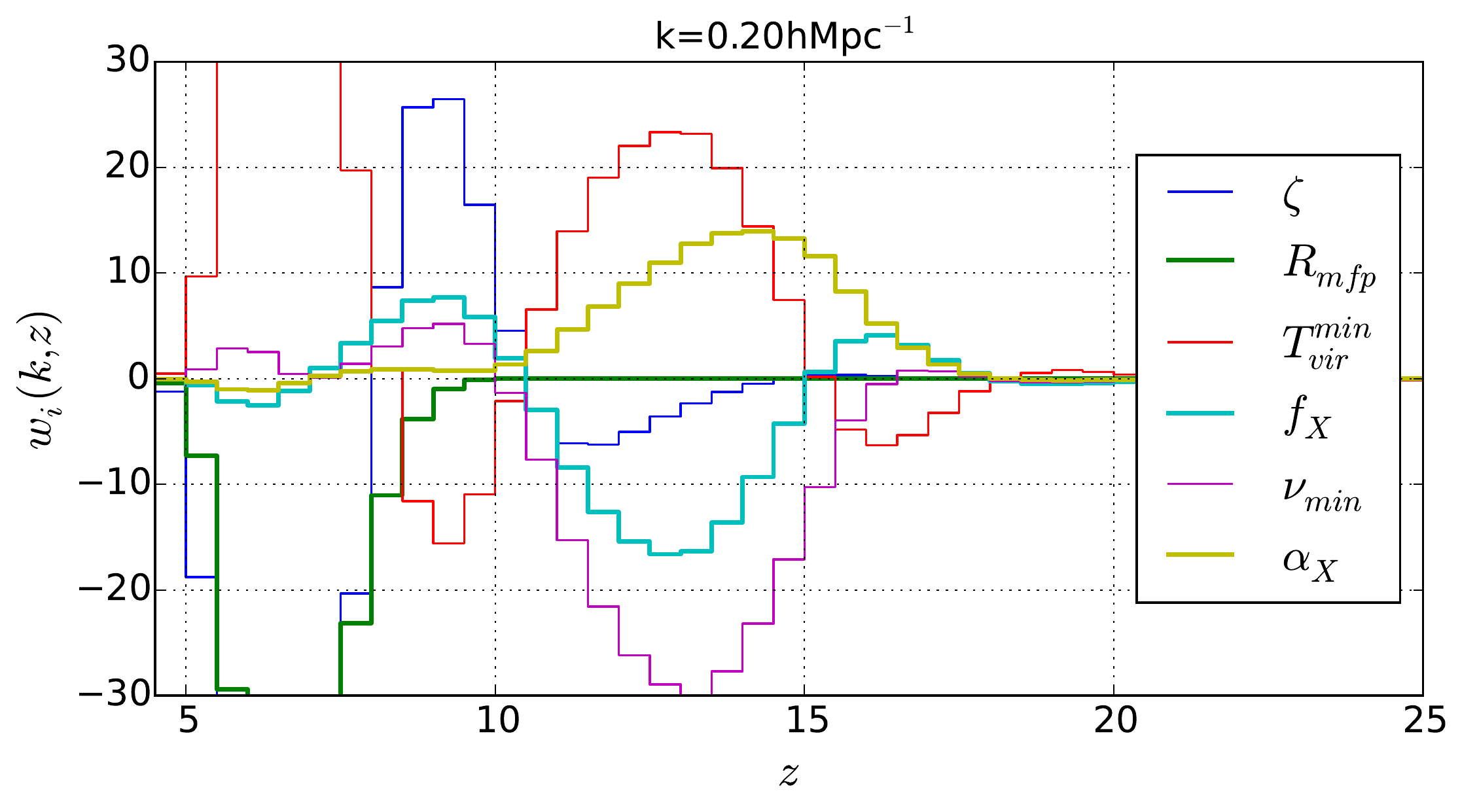}
\caption{Plotting $w_i(k,z)$ at a single cosmological Fourier mode for all of our parameters on the same panel facilitates direct comparison. Many of the parameters have similar redshift evolutions that differ by a sign, making their effects on the power spectrum degenerate.} 
\label{fig:wSingleMode}
\end{figure}

To understand specific sources of covariance, we plot $w_i$ vs. $z$ at several different comoving scales (Fig.~\ref{fig:wCompare}) and directly compare the redshift evolution between all $w_i$s at a single mode in Fig.~\ref{fig:wSingleMode}. Here we assume the thermal noise from HERA-331 and our moderate foregrounds scenario. Redshifts where $|w_i(k,z)|$ is largest indicate where our measurements will have maximal sensitivity to each $\theta_i$.

 As we might expect, the greatest information on $\zeta$ is obtained over reionization. Because $w_\zeta$ follows essentially the opposite trend of $w_T$, even out to higher redshifts, there is extensive degeneracy between the two parameters. Since $w_R$ is only significant during the end of reionization, $R_\text{mfp}$s degeneracies with $\zeta$ and $T_\text{vir}^\text{min}$ are broken with the inclusion of higher redshift observations. 
Turning to the X-ray spectral parameters; the evolution of $w_f$ and $w_\zeta$ follows very similar trends during the fall of the heating peak and reionization. While $w_\nu$ and $w_f$ have unique trends over the entire duration of reionization, their evolution is very similar during the end of the heating peak where the power spectrum is more sensitive to them. Because its impact on reionization is negligible, $\alpha_X$'s covariance with the reionization parameters will be very small. Over heating, $w_\alpha$ follows a similar and opposite trend to $w_\nu$. 

\begin{figure*}
\includegraphics[width=\textwidth]{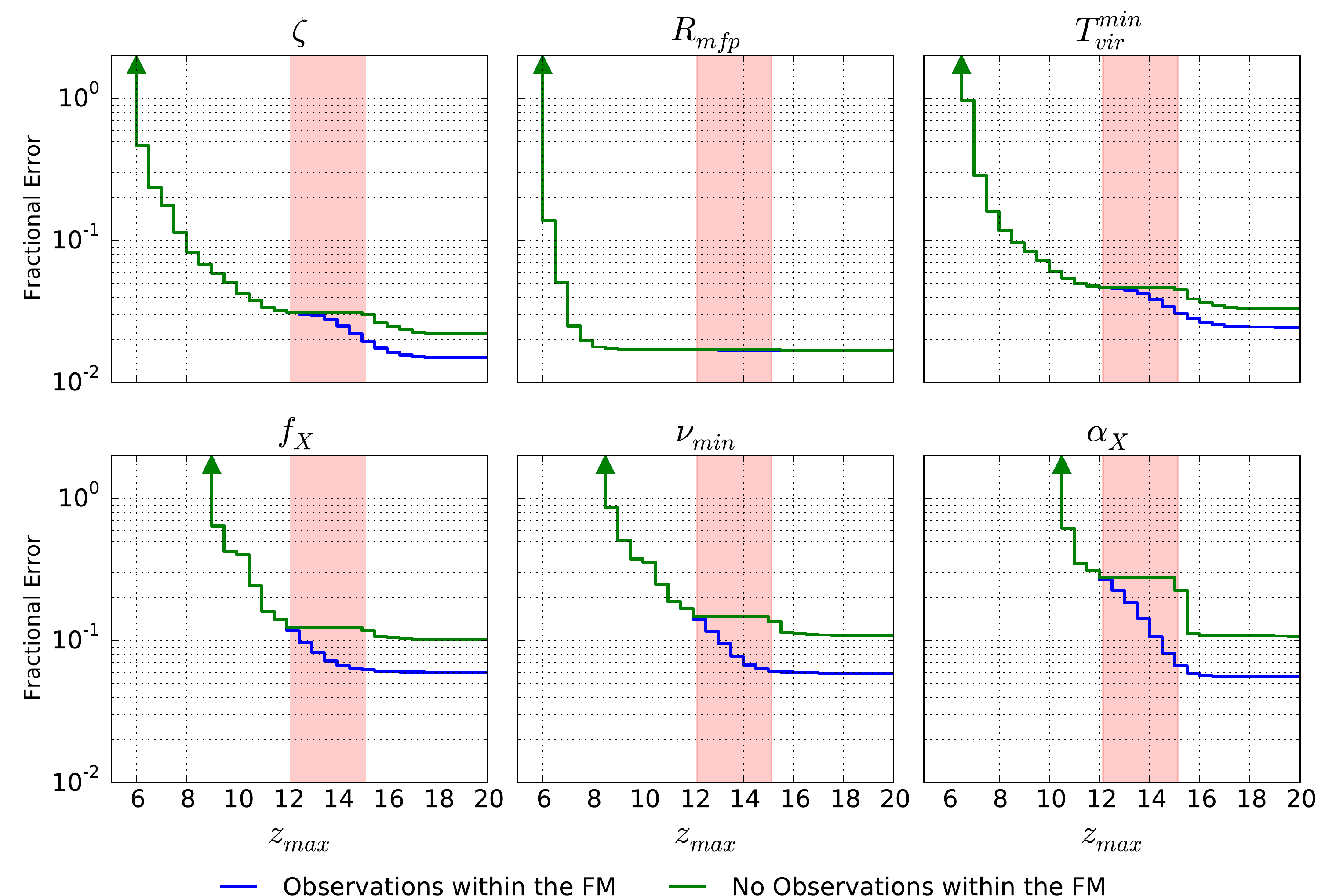}
\caption{The fractional errors on astrophysical parameters as a function of maximal observed redshift for HERA-331 with moderate foreground contamination. From low redshift measurements, $\nu_\text{min}$ and $f_X$ are constrained to within $\approx 40\%$t hough the spectral index $\alpha_X$ remains highly uncertain. Measurements at $z \gtrsim 10$ allow for $\lesssim 10\%$ limits on X-ray spectral parameters, including $\alpha_X$ and a factor of two improvement in constraints on $T_\text{vir}^\text{min}$ and reionization. Inability to observe within the FM radio-band (pink shaded region) raises the errors on heating parameters by a factor of two. The fall of the error bars with redshift bottoms out at high $z$ due to increasing thermal noise.}
\label{fig:zSigma}
\end{figure*}
\subsection{How well can Epoch of Reionization Measurements Constrain X-ray Spectral Properties?}\label{ssec:HeatFromReionization}
We now determine what constraints on X-ray spectral properties can actually be obtained by measurements of the reionization peak which, in our model, extends to roughly $z \lesssim 10$. In Fig.~\ref{fig:zSigma} we show the marginalized $1\sigma$ error bars as a function of the maximal redshift included in power spectrum observations on HERA-331 with moderate foregrounds. We see that below the onset of the reionization peak at  $z \approx 10$, the error bars are at $\approx 40\%$ for $f_x$ and $\nu_\text{min}$ while the error on $\alpha_x$ exceeds $100\%$. The latter is understandable given that $w_\alpha$ is very small below $z=10$ relative to the other parameters.

\begin{figure}
\includegraphics[width=.48\textwidth]{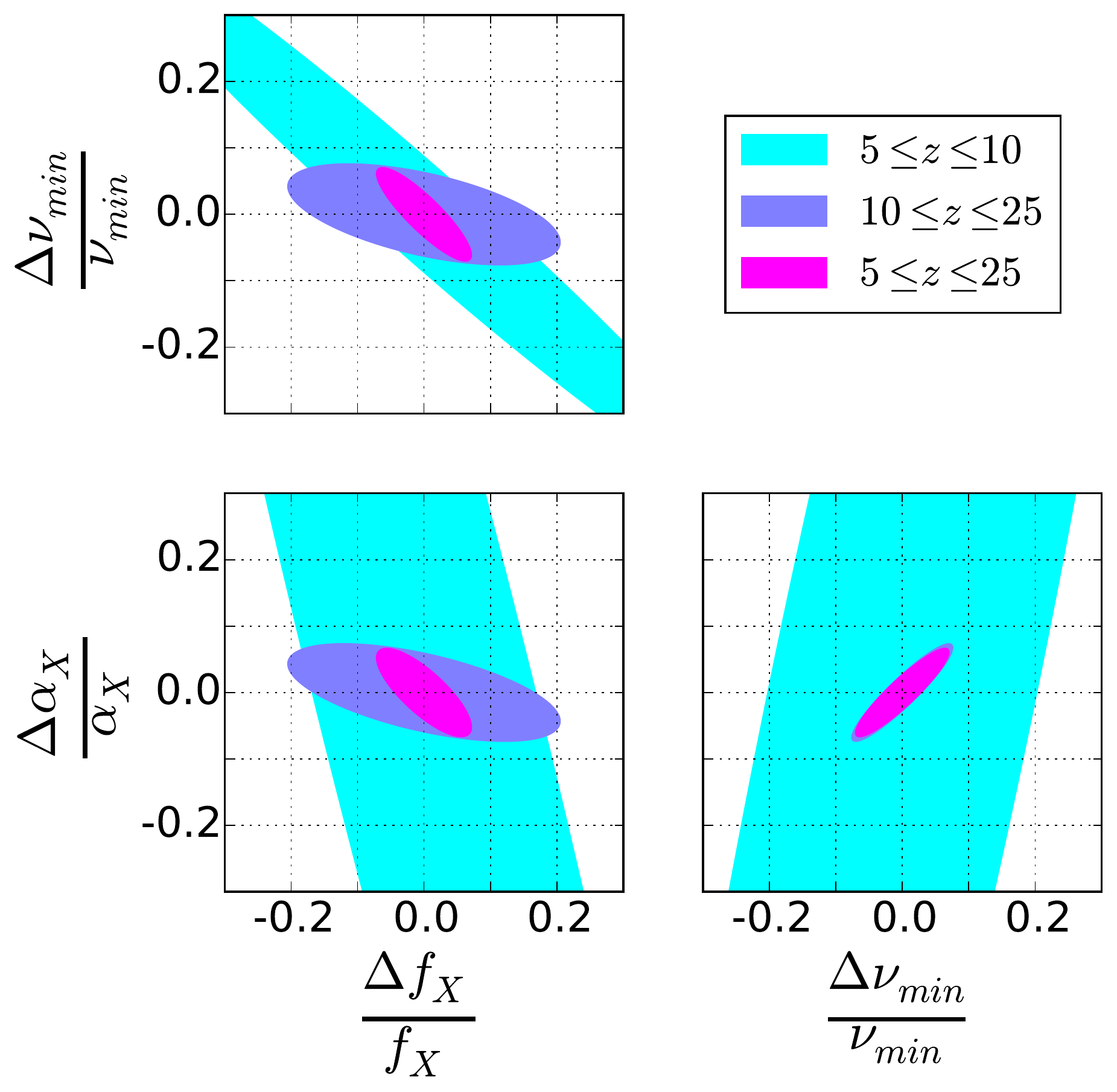}
\caption{95\% confidence regions for the X-ray spectral properties of early galaxies, marginalizing over $(\zeta,R_\text{mfp},T_\text{vir}^\text{min})$ from measurements on HERA-331 with our moderate foregrounds scenario. At low redshift $(z\leq10)$, hardly any signature of $\alpha_X$ is present, leading to large error bars on the $\alpha_X$ axis. Because $\nu_\text{min}$ and $f_X$ incur very similar changes on the power spectrum during the beginning of reionization (Fig.~\ref{fig:wCompare}), they are highly degenerate. Observations of the heating peak break these degeneracies.}
\label{fig:triangleHeating}
\end{figure}

 We delve into the source of heating uncertainty in $\nu_\text{min}$ and $f_X$ over reionization by plotting the $95\%$ confidence ellipses of our heating parameters, in Fig.~\ref{fig:triangleHeating}, from observations over reionization ($5 \lesssim z \lesssim 10$), heating ($10 \lesssim z \lesssim 25$), and both ($5 \lesssim z \lesssim 25$). We see that during reionization, there is a large negative correlation between $f_X$ and $\nu_\text{min}$ ($w_f$ in Fig.~\ref{fig:wCompare} follows very similar trends to $w_\nu$ at the end of reionization). When we fix $\nu_\text{min}$ at its fiducial value, we obtain several percent constraints on $f_X$ (instead of $40\%$) and vice versa. $\alpha_X$, covaries weakly with the other heating parameters over reionization, but has very large error bars. Eliminating it reduces the errors on $\nu_\text{min}$ and $f_X$ by a factor of two. The inclusion of heating measurements in addition to reionization, removes much of the $f_X - \nu_\text{min}$ degeneracy, bringing their fractional errors to within $\approx 6\%$. 

While signatures of heating are present in the early stages of reionization, degeneracies between heating parameters prevent precision constraints. All constraints on $\alpha_X$ come from direct measurements of the heating epoch at $z \gtrsim 15$.

\subsection{How well do Epoch of X-ray Heating measurements improve Constraints on Reionization?}\label{ssec:ReionizationFromHeat}

\begin{figure}
\includegraphics[width=.48\textwidth]{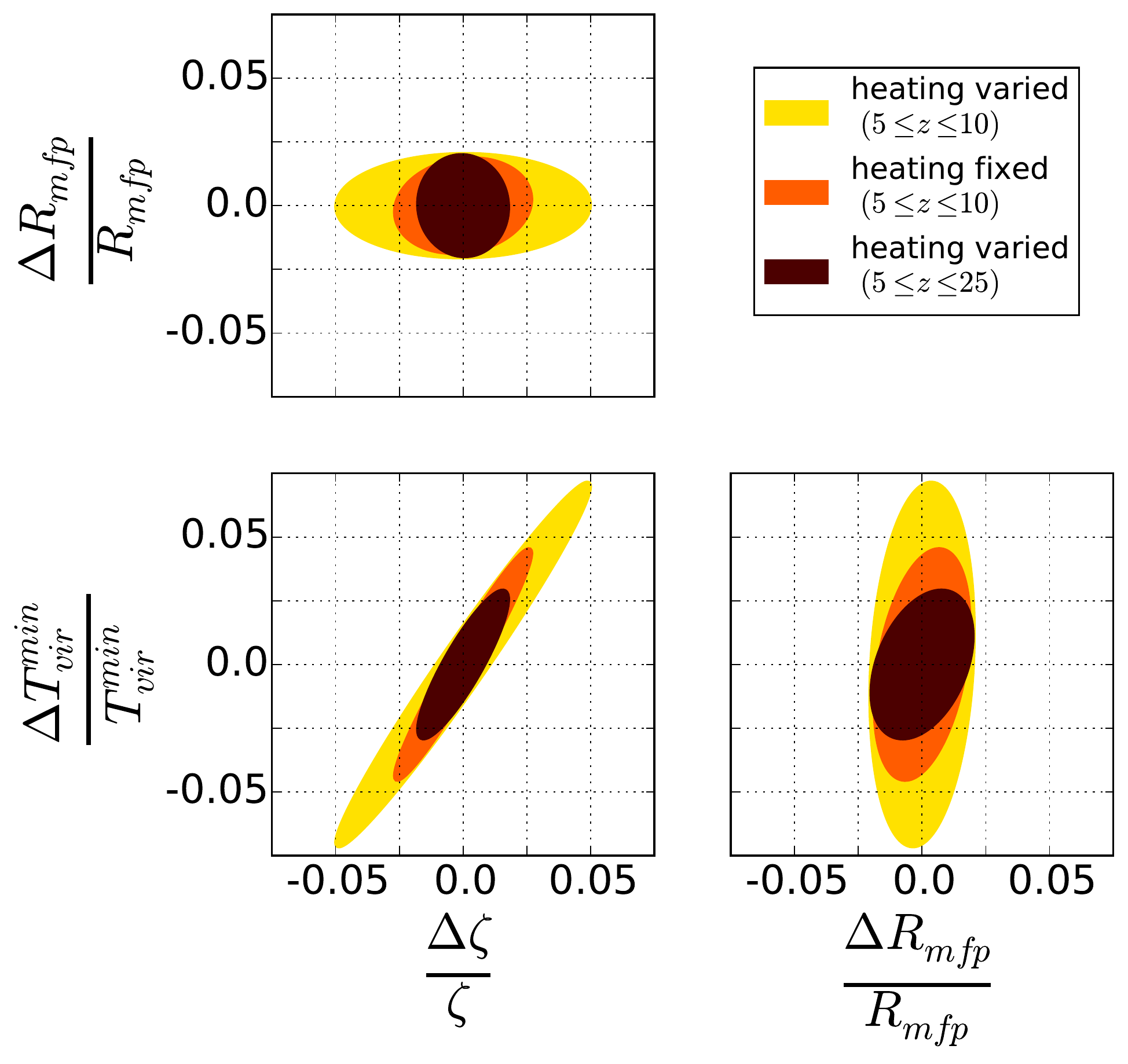}
\caption{Confidence ellipses (95\%) for $T_\text{vir}^\text{min}$ and reionization parameters. By comparing the ellipses resulting from fixing our heating history and only observing at low redshift and the ellipses resulting from marginalizing over all parameters but including heating epoch measurements, one can see that a significant fraction of the gains in reionization uncertainties at high redshift come from breaking degeneracies with heating parameters rather than the direct signatures of reionization. This also shows that not marginalizing over heating parameters leads to over-optimistic predictions of reionization uncertainties.}
\label{fig:triangleIonization}
\end{figure}

We now determine what measurements during the Epoch of X-ray heating $(z \gtrsim 10)$ can teach us about reionization. In Fig.~\ref{fig:zSigma} we see that the errors on $T_\text{vir}^\text{min}$ and $\zeta$ are reduced roughly by a factor of two when data from the heating epoch is included. Where do these improvements originate from? We know from our discussion in \S~\ref{sec:Derivatives} that information on $T_\text{vir}^\text{min}$ and $\zeta$ extends to higher redshift, providing one possible explanation. On the other hand, inspection of Fig.~\ref{fig:wSingleMode} indicates that there are also degeneracies between the reionization and heating parameters during the beginning of reionization and higher redshift measurements add information by breaking these degeneracies. We determine the impact of these two sources of information by comparing the 95\% confidence ellipses for reionization parameters derived from reionization observations in which heating parameters have been fixed and the confidence ellipses when all parameters have been marginalized over with heating observations included (Fig.~\ref{fig:triangleIonization}). While fixing heating parameters leads to a significant drop in the areas of the confidence regions, the inclusion of high redshift measurements provides additional improvements. We conclude that the improvements in reionization constraints arise through both mechanisms; breaking low redshift degeneracies with the heating parameters and obtaining additional information present in early HII bubbles. This also means that not marginalizing over heating parameters (as is done in \citet{Pober:2014} and \citet{Greig:2015a}) leads to overly-optimistic predictions of reionization constraints. We note that the heating-reionization degeneracies arise primarily at the beginning of reionization when $T_s \sim T_\text{cmb}$ (Fig.~\ref{fig:Global}). In a more efficient heating scenario, we would expect their contribution to be reduced.

\subsection{Overall Parameter Constraints}

\begin{figure*}
\includegraphics[width=\textwidth]{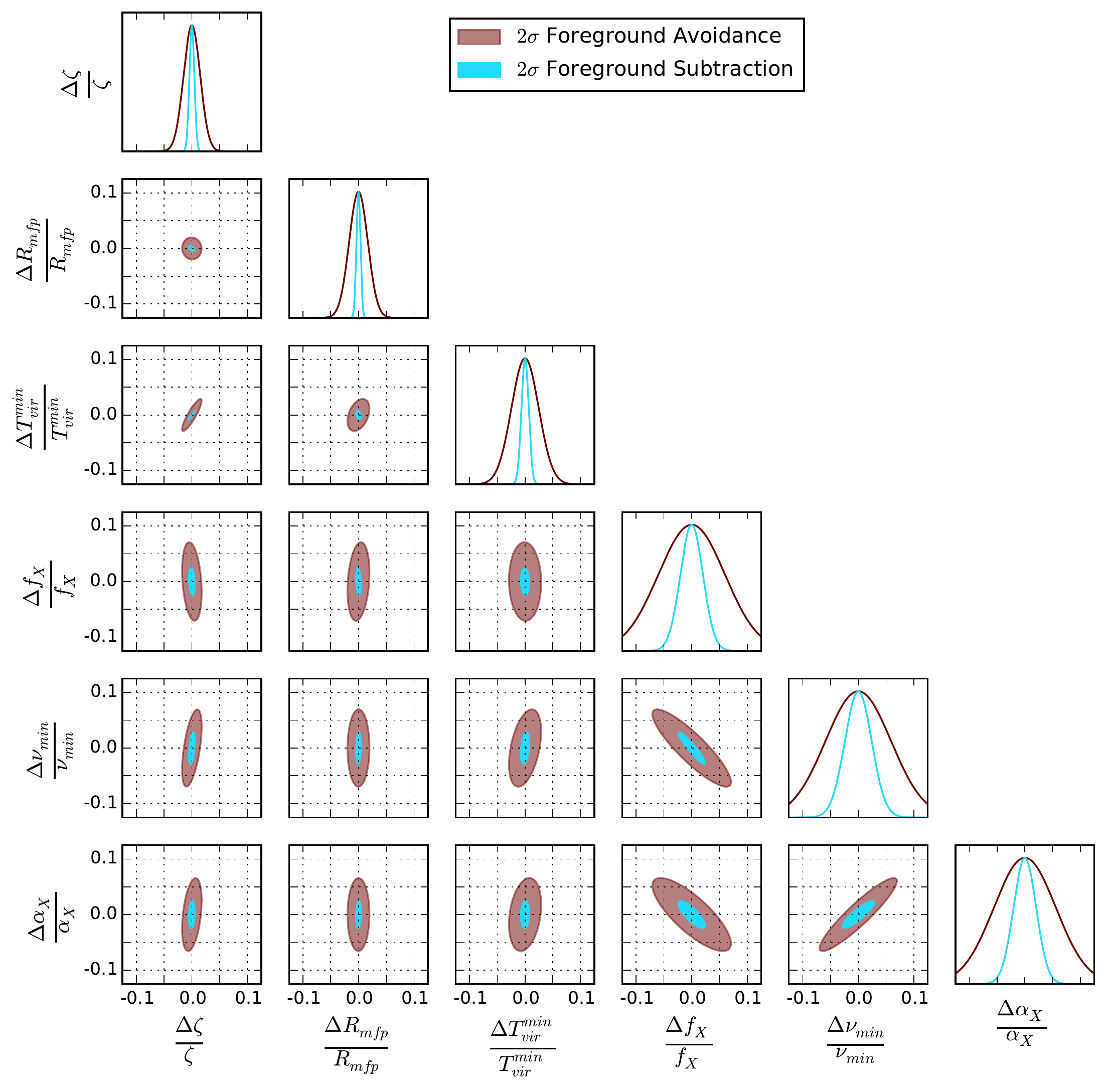}
\caption{95\% confidence ellipses for our moderate and optimistic foreground models assuming 1080 hours of drift-scan observations on HERA-331. Heating parameters tend to be highly degenerate with eachother but independent of reionization parameters. As the majority of the information heating comes from higher redshifts where thermal noise is much higher, their uncertainty regions tend to be several times larger.}\label{fig:triangle}
\end{figure*}

We observe overall degeneracies and error bars in the 95\% confidence ellipses derived from the inverse of the Fisher matrix for data between $z=5$ and $z=25$ in Fig.~\ref{fig:triangle}. Here we assume that all frequencies, including those within the FM are accessible to observations. Relatively little degeneracy exists between heating and reionization parameters, though we would not expect this to be the case for a model in which heating were delayed. 

 In Fig.~\ref{fig:zSigma} we show the $1 \sigma$ uncertainty on each astrophysical parameter as a function of the maximal redshift included in the Fisher matrix analysis. While $R_\text{mfp}$ reaches its minimal error of several percent by $z \approx 9$, the error bar on  $\zeta$ and $T_\text{vir}^\text{min}$ can be improved by nearly a factor of two by including power spectrum measurements of heating. Measurements at $z\gtrsim 10$ bring the error bars on all heating parameters below $\approx 6\%$ for HERA-331 while reionization measurements alone yield $40\%$ errors. In Fig.~\ref{fig:zSigma} we also show the effect of the FM on heating parameter error bars. Observations within the FM lead to a factor of two improvement in limits on heating, showing that the RFI environment of the observatory, within the FM, will have an important effect on the science that can be performed. It is found in \citet{EwallWice:2015} that after three hours of integration, FM is not a limiting systematic for pre-reionization observations at the Murchison Radio Observatory in Western Australia (where SKA-1 is planned). However, the amount of lower level RFI that might become a limiting obstacle after the hundreds of hours of observation necessary for a detection is unknown.  We note that the impact of the FM on our constraints is model dependent and that the heating peak for the scenario considered in this work occurs at $z \approx 15$, right in the middle of the FM (88-108\,MHz). Models with a different fiducial $f_X$ would produce a power spectrum peak at higher or lower redshifts, making the scenario that we consider here pessimistic with respect to the FM's impact on science.

We generalize our discussion to additional instrument and foreground scenarios discussed in \S~\ref{sec:MethodsNoise} by displaying forecasts of $1\sigma$ fractional parameter uncertainties in Table~\ref{tab:errors}. In the pessimistic scenario, the baselines that are not instantaneously redundant are never added coherently, leading to a very significant reduction in the SKA's sensitivity and preventing it from placing any significant limits on X-ray heating. In the foreground avoidance scenario, SKA-1 and HERA-127 both place $\approx 10-15$\% constraints on heating parameters while HERA-331 obtains $\approx 6$\% error bars. Should we obtain sufficient characterization of foregrounds as to allow us to subtract them and work within the wedge, then several percent to sub-percent constraints are possible with HERA-331 and the SKA which is far beyond the the modeling uncertainties in our semi-numerical framework.

\section{Conclusions}\label{sec:Conclusions}
	Measurements of the 21\,cm power spectrum during the EoR are poised to put significant limits on the properties of the UV sources that ionized the IGM. At higher redshifts, the power spectrum is heavily influenced by X-rays from accretion onto the first stellar mass black holes and ISM heated by the first supernovae. While reionization has a number of complementary probes, observations of the 21\,cm global signal and power spectrum at these higher redshifts provide us with what is likely the only means of obtaining detailed knowledge on the earliest X-ray sources and their impact on future generations of galaxies. 
	
	In this paper we have used the Fisher matrix formalism and semi-numerical simulations to take a first step in quantifying the accuracy with which upcoming experiments will constrain the properties of the first X-ray sources. Our analysis also aims to understand what additional constraints on reionization parameters exist at higher redshift when the spin temperature calculation is included and whether higher redshift observations might break degeneracies between reionization parameters such as the degeneracy between $\zeta$ and $T_\text{vir}^\text{min}$.

	We have found that the detectable impact of the ionization efficiency is manifested in the form of early HII holes around IGM hotspots and that the inclusion of the spin temperature calculation and the additional heating parameters increases our uncertainty of reionization parameters through new degeneracies. Observations of the heating epoch reduce the errors on reionization parameters by a factor of two by accessing the information in early HII bubbles and breaking degeneracies between ionization and heating during the beginning of reionization. Since previous works ignored the degeneracies of reionization with heating the predictions in these works are therefor optimistic by a factor of $\sim 2$. 
	
	Though heating does have an effect on the reionization power spectrum as noted by \citet{Mesinger:2013} and \citet{Fialkov:2014} and clearly visible in the non-zero derivatives at $z=9$ in Fig.~\ref{fig:Derivatives} in our paper, the effects of different heating parameters are highly degenerate leading to $\gtrsim40\%$ fractional error bars unless higher redshift observations are folded in. Information on the detailed spectral properties of the sources, which would enable us to discriminate between hot ISM or HMXB heating as well as precision constrains on other heating parameters will likely require power spectrum measurements at $z \gtrsim 10$. In the model we study here, HERA and SKA-low will be able to place $\approx 6-10\%$ constrains on heating parameters even if observations in the FM band are not possible.

	In this analysis, we have chosen to examine a single model in a large allowed parameter space. We do not think our predictions will change in different models by more than an order of magnitude based on trends that are well documented in the literature. It is shown in \citet{Pober:2014} that HERA is capable of detecting the reionization peak over several orders of magnitude in $T_\text{vir}^\text{min}$ and a wide range of $\zeta$ and $R_\text{mfp}$ values. The height of the heating peak is constant through 3-4 orders of magnitude in $f_X$ \citep{Mesinger:2014} while the redshift of the peak remains approximately between $10$ and $20$ \citep{Pacucci:2014}, hence the SNR on heating should not vary by more than an order of magnitude. An $\alpha_X$ that is larger than our fiducial value by a factor of $2-3$ is shown, in \citet{Pacucci:2014}, to boost the amplitude of the heating peak by a factor of $\approx 2-3$ which we would expect to improve our constraints at a similar level. A much harder X-ray spectrum such as that discussed in \citet{Fialkov:2014} leads to a reduction in the power spectrum amplitude by a factor of $\approx 2$ but also shift the peak to lower redshifts. We would expect the combination of these effects to give results within an order of magnitude of those presented here.
	
	In adopting the Fisher matrix technique, we have assumed that our likelihood function is Gaussian. Though this is a reasonable approximation about the ML point for small error bars, a more robust approach would be an MCMC calculation such as that presented in \citet{Greig:2015a}. It is for this reason that we do not give projections for the sensitivity of current arrays since projections of large error bars using the Fisher matrix are not self consistent. As of now, calculations of heating are not sufficiently rapid to allow for MCMC sampling of the likelihood function. Speeding up the heating calculation for suitability in MCMC is the subject of ongoing work.

	Finally, with the exception of RFI, we have not attempted to directly address the fact that known systematic obstacles to 21\,cm observing, such as the ionosphere, foreground brightness, and the increasing extent of the primary beam, become worse at lower frequencies and may pose challenges in addition to the the ones we address: namely RFI and increased thermal noise. Observations in \citet{EwallWice:2015} find that while ionospheric refraction does not appear to impact the level of foreground leakage beyond the wedge, the brighter foregrounds extended primary beam heighten the severity of any uncalibrated structure in the instrumental bandpass.

Our results indicate that precision measurements of the first high energy galactic processes can be expected from the upcoming generation of power spectrum experiments, provided that they exploit the information in the redshifts typically considered to precede reionization.

\begin{table*}
\centering
\begin{tabular}{c|ccc|ccc|ccc} 
 \multirow{2}{*}{{\bf Parameters}} & \multicolumn{3}{c|}{{\bf Pessimistic Foregrounds}} & \multicolumn{3}{c|}{{\bf Moderate Foregrounds}} & \multicolumn{3}{c}{{\bf Optimistic Foregrounds}} \\ 
 &  HERA-127 & HERA-331 & SKA-1 & HERA-127 & HERA-331 & SKA-1 & HERA-127 & HERA-331 & SKA-1  \\ \hline
$\Delta \zeta/\zeta^\text{fid}$ & 0.04 & 0.02 & 0.16 & 0.03 & 0.01 & 0.03 & 0.01 & 0.004 & 0.004 \\
$\Delta R_\text{mfp}/R_\text{mfp}^\text{fid}$ & 0.03 & 0.02 & 0.10 & 0.02 & 0.02 & 0.04 & 0.01 & 0.004 & 0.003 \\
$\Delta T_\text{vir}^\text{min}/T_\text{vir}^\text{min,fid}$ & 0.05 & 0.03 & 0.23 & 0.04 & 0.02 & 0.05 & 0.02 & 0.007 & 0.006 \\
$\Delta f_X/f_X$ & 0.18 & 0.07 & 0.79 & 0.15 & 0.06 & 0.13 & 0.06 & 0.020 & 0.020 \\
$\Delta \nu_\text{min}/\nu^\text{fid}_\text{min}$ & 0.19 & 0.07 & 0.79 & 0.15 & 0.06 & 0.12 & 0.07 & 0.024 & 0.020  \\
$\Delta \alpha_X/\alpha^\text{fid}_X$ & 0.15 & 0.07 & 0.75 & 0.13 & 0.06 & 0.13 & 0.06 & 0.020 & 0.023
\end{tabular}
\caption{The $1\sigma$ error forecasts for reionization and heating parameters on the instruments studied in this paper assuming $1080$ hours of drift-scan observations on $\Delta z =0.5$ co-eval bands over all redshifts between $5$ and $25$.}
\label{tab:errors}
\end{table*}

\section*{Agknowledgements}
A.E.W. and J.H. thank the Massachusetts Institute of Technology School of Science for support and also acknowledge support from the MRI grant NSF AST-0821321. A.E.W. acknowledges support from the National Science Foundation Graduate Research Fellowship under Grant No. 1122374. A.M. acknowledges support from the European Research Council (ERC) under the European Unions Horizon 2020 research and innovation program (grant agreement No 638809 AIDA). J.S.D. gratefully acknowledges support from the Berkeley Center for Cosmological Physics. A.L. acknowledges support for this work by NASA through Hubble Fellowship grant \#HST-HF2- 51363.001-A awarded by the Space Telescope Science Institute, which is operated by the Association of Universities for Research in Astronomy, Inc., for NASA, under contract NAS5-26555. J.C.P acknowledges NSF support through award \#1302774

\bibliographystyle{mnras}
\bibliography{ms}

\newcommand{\noop}[1]{}
\begin{thebibliography}{}
\makeatletter
\relax
\def\mn@urlcharsother{\let\do\@makeother \do\$\do\&\do\#\do\^\do\_\do\%\do\~}
\def\mn@doi{\begingroup\mn@urlcharsother \@ifnextchar [ {\mn@doi@}
  {\mn@doi@[]}}
\def\mn@doi@[#1]#2{\def\@tempa{#1}\ifx\@tempa\@empty \href
  {http://dx.doi.org/#2} {doi:#2}\else \href {http://dx.doi.org/#2} {#1}\fi
  \endgroup}
\def\mn@eprint#1#2{\mn@eprint@#1:#2::\@nil}
\def\mn@eprint@arXiv#1{\href {http://arxiv.org/abs/#1} {{\tt arXiv:#1}}}
\def\mn@eprint@dblp#1{\href {http://dblp.uni-trier.de/rec/bibtex/#1.xml}
  {dblp:#1}}
\def\mn@eprint@#1:#2:#3:#4\@nil{\def\@tempa {#1}\def\@tempb {#2}\def\@tempc
  {#3}\ifx \@tempc \@empty \let \@tempc \@tempb \let \@tempb \@tempa \fi \ifx
  \@tempb \@empty \def\@tempb {arXiv}\fi \@ifundefined
  {mn@eprint@\@tempb}{\@tempb:\@tempc}{\expandafter \expandafter \csname
  mn@eprint@\@tempb\endcsname \expandafter{\@tempc}}}

\bibitem[\protect\citeauthoryear{{Ali} et~al.,}{{Ali} et~al.}{2015}]{Ali:2015}
{Ali} Z.~S.,  et~al., 2015, \mn@doi [\apj] {10.1088/0004-637X/809/1/61}, \href
  {http://adsabs.harvard.edu/abs/2015ApJ...809...61A} {809, 61}

\bibitem[\protect\citeauthoryear{{Baek}, {Semelin}, {Di Matteo}, {Revaz}  \&
  {Combes}}{{Baek} et~al.}{2010}]{Baek:2010}
{Baek} S.,  {Semelin} B.,  {Di Matteo} P.,  {Revaz} Y.,   {Combes} F.,  2010,
  \mn@doi [\aap] {10.1051/0004-6361/201014347}, \href
  {http://adsabs.harvard.edu/abs/2010A%26A...523A...4B} {523, A4}

\bibitem[\protect\citeauthoryear{{Barkana} \& {Loeb}}{{Barkana} \&
  {Loeb}}{2004}]{Barkana:2004}
{Barkana} R.,  {Loeb} A.,  2004, \mn@doi [\apj] {10.1086/421079}, \href
  {http://adsabs.harvard.edu/abs/2004ApJ...609..474B} {609, 474}

\bibitem[\protect\citeauthoryear{{Basu-Zych} et~al.,}{{Basu-Zych}
  et~al.}{2013}]{Basu-Zych:2013}
{Basu-Zych} A.~R.,  et~al., 2013, \mn@doi [\apj] {10.1088/0004-637X/762/1/45},
  \href {http://adsabs.harvard.edu/abs/2013ApJ...762...45B} {762, 45}

\bibitem[\protect\citeauthoryear{{Bernardi}, {McQuinn}  \&
  {Greenhill}}{{Bernardi} et~al.}{2015}]{Bernardi:2015}
{Bernardi} G.,  {McQuinn} M.,   {Greenhill} L.~J.,  2015, \mn@doi [\apj]
  {10.1088/0004-637X/799/1/90}, \href
  {http://adsabs.harvard.edu/abs/2015ApJ...799...90B} {799, 90}

\bibitem[\protect\citeauthoryear{{Bowman} \& {Rogers}}{{Bowman} \&
  {Rogers}}{2010}]{Bowman:2010}
{Bowman} J.~D.,  {Rogers} A.~E.~E.,  2010, \mn@doi [\nat]
  {10.1038/nature09601}, \href
  {http://adsabs.harvard.edu/abs/2010Natur.468..796B} {468, 796}

\bibitem[\protect\citeauthoryear{{Bunn}, {Scott}  \& {White}}{{Bunn}
  et~al.}{1995}]{Bunn:1995}
{Bunn} E.~F.,  {Scott} D.,   {White} M.,  1995, \mn@doi [\apjl]
  {10.1086/187776}, \href {http://adsabs.harvard.edu/abs/1995ApJ...441L...9B}
  {441, L9}

\bibitem[\protect\citeauthoryear{{Burns} et~al.,}{{Burns}
  et~al.}{2012}]{Burns:2012}
{Burns} J.~O.,  et~al., 2012, \mn@doi [Advances in Space Research]
  {10.1016/j.asr.2011.10.014}, \href
  {http://adsabs.harvard.edu/abs/2012AdSpR..49..433B} {49, 433}

\bibitem[\protect\citeauthoryear{{Christian} \& {Loeb}}{{Christian} \&
  {Loeb}}{2013}]{Christian:2013}
{Christian} P.,  {Loeb} A.,  2013, \mn@doi [\jcap]
  {10.1088/1475-7516/2013/09/014}, \href
  {http://adsabs.harvard.edu/abs/2013JCAP...09..014C} {9, 14}

\bibitem[\protect\citeauthoryear{{Cowie}, {Barger}  \& {Hasinger}}{{Cowie}
  et~al.}{2012}]{Cowie:2012}
{Cowie} L.~L.,  {Barger} A.~J.,   {Hasinger} G.,  2012, \mn@doi [\apj]
  {10.1088/0004-637X/748/1/50}, \href
  {http://adsabs.harvard.edu/abs/2012ApJ...748...50C} {748, 50}

\bibitem[\protect\citeauthoryear{{Datta}, {Bowman}  \& {Carilli}}{{Datta}
  et~al.}{2010}]{Datta:2010}
{Datta} A.,  {Bowman} J.~D.,   {Carilli} C.~L.,  2010, \mn@doi [\apj]
  {10.1088/0004-637X/724/1/526}, \href
  {http://adsabs.harvard.edu/abs/2010ApJ...724..526D} {724, 526}

\bibitem[\protect\citeauthoryear{Dewdney}{Dewdney}{2013}]{Dewdney:2013}
Dewdney P.,  2013, Technical report, {SKA1 SYSTEM BASELINE DESIGN}

\bibitem[\protect\citeauthoryear{{Dillon}, {Liu}  \& {Tegmark}}{{Dillon}
  et~al.}{2013}]{Dillon:2013}
{Dillon} J.~S.,  {Liu} A.,   {Tegmark} M.,  2013, \mn@doi [\prd]
  {10.1103/PhysRevD.87.043005}, \href
  {http://adsabs.harvard.edu/abs/2013PhRvD..87d3005D} {87, 043005}

\bibitem[\protect\citeauthoryear{{Dillon} et~al.,}{{Dillon}
  et~al.}{2014}]{Dillon:2014}
{Dillon} J.~S.,  et~al., 2014, \mn@doi [\prd] {10.1103/PhysRevD.89.023002},
  \href {http://adsabs.harvard.edu/abs/2014PhRvD..89b3002D} {89, 023002}

\bibitem[\protect\citeauthoryear{{Dillon} et~al.,}{{Dillon}
  et~al.}{2015}]{Dillon:2015b}
{Dillon} J.~S.,  et~al., 2015, \mn@doi [\prd] {10.1103/PhysRevD.91.123011},
  \href {http://adsabs.harvard.edu/abs/2015PhRvD..91l3011D} {91, 123011}

\bibitem[\protect\citeauthoryear{{Ewall-Wice} et~al.,}{{Ewall-Wice}
  et~al.}{2015}]{EwallWice:2015}
{Ewall-Wice} A.,  et~al., 2015, {Submitted to MNRAS}

\bibitem[\protect\citeauthoryear{{Fialkov}, {Barkana}  \& {Visbal}}{{Fialkov}
  et~al.}{2014}]{Fialkov:2014}
{Fialkov} A.,  {Barkana} R.,   {Visbal} E.,  2014, \mn@doi [\nat]
  {10.1038/nature12999}, \href
  {http://adsabs.harvard.edu/abs/2014Natur.506..197F} {506, 197}

\bibitem[\protect\citeauthoryear{{Field}}{{Field}}{1959}]{Field:1959}
{Field} G.~B.,  1959, \mn@doi [\apj] {10.1086/146653}, \href
  {http://adsabs.harvard.edu/abs/1959ApJ...129..536F} {129, 536}

\bibitem[\protect\citeauthoryear{Fisher}{Fisher}{1935}]{Fisher:1935}
Fisher R.~A.,  1935, Journal of the Royal Statistical Society, 98, pp. 39

\bibitem[\protect\citeauthoryear{{Fixsen} et~al.,}{{Fixsen}
  et~al.}{2011}]{Fixsen:2011}
{Fixsen} D.~J.,  et~al., 2011, \mn@doi [\apj] {10.1088/0004-637X/734/1/5},
  \href {http://adsabs.harvard.edu/abs/2011ApJ...734....5F} {734, 5}

\bibitem[\protect\citeauthoryear{{Fragos} et~al.,}{{Fragos}
  et~al.}{2013}]{Fragos:2013}
{Fragos} T.,  et~al., 2013, \mn@doi [\apj] {10.1088/0004-637X/764/1/41}, \href
  {http://adsabs.harvard.edu/abs/2013ApJ...764...41F} {764, 41}

\bibitem[\protect\citeauthoryear{{Furlanetto}}{{Furlanetto}}{2006}]{Furlanetto:2006Global}
{Furlanetto} S.~R.,  2006, \mn@doi [\mnras] {10.1111/j.1365-2966.2006.10725.x},
  \href {http://adsabs.harvard.edu/abs/2006MNRAS.371..867F} {371, 867}

\bibitem[\protect\citeauthoryear{{Furlanetto} \& {Stoever}}{{Furlanetto} \&
  {Stoever}}{2010}]{Furlanetto:2010}
{Furlanetto} S.~R.,  {Stoever} S.~J.,  2010, \mn@doi [\mnras]
  {10.1111/j.1365-2966.2010.16401.x}, \href
  {http://adsabs.harvard.edu/abs/2010MNRAS.404.1869F} {404, 1869}

\bibitem[\protect\citeauthoryear{{Furlanetto}, {Zaldarriaga}  \&
  {Hernquist}}{{Furlanetto} et~al.}{2004}]{Furlanetto:2004}
{Furlanetto} S.~R.,  {Zaldarriaga} M.,   {Hernquist} L.,  2004, \mn@doi [\apj]
  {10.1086/423025}, \href {http://adsabs.harvard.edu/abs/2004ApJ...613....1F}
  {613, 1}

\bibitem[\protect\citeauthoryear{{Furlanetto}, {Oh}  \& {Briggs}}{{Furlanetto}
  et~al.}{2006}]{Furlanetto:2006Review}
{Furlanetto} S.~R.,  {Oh} S.~P.,   {Briggs} F.~H.,  2006, \mn@doi [\physrep]
  {10.1016/j.physrep.2006.08.002}, \href
  {http://adsabs.harvard.edu/abs/2006PhR...433..181F} {433, 181}

\bibitem[\protect\citeauthoryear{{Greenhill} \& {Bernardi}}{{Greenhill} \&
  {Bernardi}}{2012}]{GreenHill:2012}
{Greenhill} L.~J.,  {Bernardi} G.,  2012, preprint, \href
  {http://adsabs.harvard.edu/abs/2012arXiv1201.1700G} {} (\mn@eprint {arXiv}
  {1201.1700})

\bibitem[\protect\citeauthoryear{{Greig} \& {Mesinger}}{{Greig} \&
  {Mesinger}}{2015}]{Greig:2015a}
{Greig} B.,  {Mesinger} A.,  2015, preprint, \href
  {http://adsabs.harvard.edu/abs/2015arXiv150106576G} {} (\mn@eprint {arXiv}
  {1501.06576})

\bibitem[\protect\citeauthoryear{{Greig}, {Mesinger}  \& {Pober}}{{Greig}
  et~al.}{2015a}]{Greig:2015b}
{Greig} B.,  {Mesinger} A.,   {Pober} J.~C.,  2015a, preprint, \href
  {http://adsabs.harvard.edu/abs/2015arXiv150902158G} {} (\mn@eprint {arXiv}
  {1509.02158})

\bibitem[\protect\citeauthoryear{{Greig}, {Mesinger}  \& {Koopmans}}{{Greig}
  et~al.}{2015b}]{Greig:2015c}
{Greig} B.,  {Mesinger} A.,   {Koopmans} L.~V.~E.,  2015b, preprint, \href
  {http://adsabs.harvard.edu/abs/2015arXiv150903312G} {} (\mn@eprint {arXiv}
  {1509.03312})

\bibitem[\protect\citeauthoryear{{Haiman}, {Thoul}  \& {Loeb}}{{Haiman}
  et~al.}{1996a}]{Haiman:1996a}
{Haiman} Z.,  {Thoul} A.~A.,   {Loeb} A.,  1996a, \mn@doi [\apj]
  {10.1086/177343}, \href {http://adsabs.harvard.edu/abs/1996ApJ...464..523H}
  {464, 523}

\bibitem[\protect\citeauthoryear{{Haiman}, {Thoul}  \& {Loeb}}{{Haiman}
  et~al.}{1996b}]{Haiman:1996b}
{Haiman} Z.,  {Thoul} A.~A.,   {Loeb} A.,  1996b, \mn@doi [\apj]
  {10.1086/177343}, \href {http://adsabs.harvard.edu/abs/1996ApJ...464..523H}
  {464, 523}

\bibitem[\protect\citeauthoryear{{Haiman}, {Rees}  \& {Loeb}}{{Haiman}
  et~al.}{1997}]{Haiman:1997}
{Haiman} Z.,  {Rees} M.~J.,   {Loeb} A.,  1997, \apj, \href
  {http://adsabs.harvard.edu/abs/1997ApJ...476..458H} {476, 458}

\bibitem[\protect\citeauthoryear{{Hazelton}, {Morales}  \&
  {Sullivan}}{{Hazelton} et~al.}{2013}]{Hazelton:2013}
{Hazelton} B.~J.,  {Morales} M.~F.,   {Sullivan} I.~S.,  2013, \mn@doi [\apj]
  {10.1088/0004-637X/770/2/156}, \href
  {http://adsabs.harvard.edu/abs/2013ApJ...770..156H} {770, 156}

\bibitem[\protect\citeauthoryear{{Jacobs} et~al.,}{{Jacobs}
  et~al.}{2015}]{Jacobs:2015}
{Jacobs} D.~C.,  et~al., 2015, \mn@doi [\apj] {10.1088/0004-637X/801/1/51},
  \href {http://adsabs.harvard.edu/abs/2015ApJ...801...51J} {801, 51}

\bibitem[\protect\citeauthoryear{{Liu} \& {Tegmark}}{{Liu} \&
  {Tegmark}}{2011}]{Liu:2011}
{Liu} A.,  {Tegmark} M.,  2011, \mn@doi [\prd] {10.1103/PhysRevD.83.103006},
  \href {http://adsabs.harvard.edu/abs/2011PhRvD..83j3006L} {83, 103006}

\bibitem[\protect\citeauthoryear{{Liu}, {Parsons}  \& {Trott}}{{Liu}
  et~al.}{2014a}]{Liu:2014a}
{Liu} A.,  {Parsons} A.~R.,   {Trott} C.~M.,  2014a, \mn@doi [\prd]
  {10.1103/PhysRevD.90.023018}, \href
  {http://adsabs.harvard.edu/abs/2014PhRvD..90b3018L} {90, 023018}

\bibitem[\protect\citeauthoryear{{Liu}, {Parsons}  \& {Trott}}{{Liu}
  et~al.}{2014b}]{Liu:2014b}
{Liu} A.,  {Parsons} A.~R.,   {Trott} C.~M.,  2014b, \mn@doi [\prd]
  {10.1103/PhysRevD.90.023019}, \href
  {http://adsabs.harvard.edu/abs/2014PhRvD..90b3019L} {90, 023019}

\bibitem[\protect\citeauthoryear{{Mao}, {Tegmark}, {McQuinn}, {Zaldarriaga}  \&
  {Zahn}}{{Mao} et~al.}{2008}]{Mao:2008}
{Mao} Y.,  {Tegmark} M.,  {McQuinn} M.,  {Zaldarriaga} M.,   {Zahn} O.,  2008,
  \mn@doi [\prd] {10.1103/PhysRevD.78.023529}, \href
  {http://adsabs.harvard.edu/abs/2008PhRvD..78b3529M} {78, 023529}

\bibitem[\protect\citeauthoryear{{Mesinger} \& {Dijkstra}}{{Mesinger} \&
  {Dijkstra}}{2008}]{Mesinger:2008}
{Mesinger} A.,  {Dijkstra} M.,  2008, \mn@doi [\mnras]
  {10.1111/j.1365-2966.2008.13776.x}, \href
  {http://adsabs.harvard.edu/abs/2008MNRAS.390.1071M} {390, 1071}

\bibitem[\protect\citeauthoryear{{Mesinger} \& {Furlanetto}}{{Mesinger} \&
  {Furlanetto}}{2007}]{Mesinger:2007}
{Mesinger} A.,  {Furlanetto} S.,  2007, \mn@doi [\apj] {10.1086/521806}, \href
  {http://adsabs.harvard.edu/abs/2007ApJ...669..663M} {669, 663}

\bibitem[\protect\citeauthoryear{{Mesinger}, {Furlanetto}  \& {Cen}}{{Mesinger}
  et~al.}{2011}]{Mesinger:2011}
{Mesinger} A.,  {Furlanetto} S.,   {Cen} R.,  2011, \mn@doi [\mnras]
  {10.1111/j.1365-2966.2010.17731.x}, \href
  {http://adsabs.harvard.edu/abs/2011MNRAS.411..955M} {411, 955}

\bibitem[\protect\citeauthoryear{{Mesinger}, {McQuinn}  \&
  {Spergel}}{{Mesinger} et~al.}{2012}]{Mesinger:2012}
{Mesinger} A.,  {McQuinn} M.,   {Spergel} D.~N.,  2012, \mn@doi [\mnras]
  {10.1111/j.1365-2966.2012.20713.x}, \href
  {http://adsabs.harvard.edu/abs/2012MNRAS.422.1403M} {422, 1403}

\bibitem[\protect\citeauthoryear{{Mesinger}, {Ferrara}  \&
  {Spiegel}}{{Mesinger} et~al.}{2013}]{Mesinger:2013}
{Mesinger} A.,  {Ferrara} A.,   {Spiegel} D.~S.,  2013, \mn@doi [\mnras]
  {10.1093/mnras/stt198}, \href
  {http://adsabs.harvard.edu/abs/2013MNRAS.431..621M} {431, 621}

\bibitem[\protect\citeauthoryear{{Mesinger}, {Ewall-Wice}  \&
  {Hewitt}}{{Mesinger} et~al.}{2014}]{Mesinger:2014}
{Mesinger} A.,  {Ewall-Wice} A.,   {Hewitt} J.,  2014, \mn@doi [\mnras]
  {10.1093/mnras/stu125}, \href
  {http://adsabs.harvard.edu/abs/2014MNRAS.439.3262M} {439, 3262}

\bibitem[\protect\citeauthoryear{{Mineo}, {Gilfanov}  \& {Sunyaev}}{{Mineo}
  et~al.}{2012a}]{Mineo:2012a}
{Mineo} S.,  {Gilfanov} M.,   {Sunyaev} R.,  2012a, \mn@doi [\mnras]
  {10.1111/j.1365-2966.2011.19862.x}, \href
  {http://adsabs.harvard.edu/abs/2012MNRAS.419.2095M} {419, 2095}

\bibitem[\protect\citeauthoryear{{Mineo}, {Gilfanov}  \& {Sunyaev}}{{Mineo}
  et~al.}{2012b}]{Mineo:2012b}
{Mineo} S.,  {Gilfanov} M.,   {Sunyaev} R.,  2012b, \mn@doi [\mnras]
  {10.1111/j.1365-2966.2012.21831.x}, \href
  {http://adsabs.harvard.edu/abs/2012MNRAS.426.1870M} {426, 1870}

\bibitem[\protect\citeauthoryear{{Mirabel}, {Dijkstra}, {Laurent}, {Loeb}  \&
  {Pritchard}}{{Mirabel} et~al.}{2011}]{Mirabel:2011}
{Mirabel} I.~F.,  {Dijkstra} M.,  {Laurent} P.,  {Loeb} A.,   {Pritchard}
  J.~R.,  2011, \mn@doi [\aap] {10.1051/0004-6361/201016357}, \href
  {http://adsabs.harvard.edu/abs/2011A%26A...528A.149M} {528, A149}

\bibitem[\protect\citeauthoryear{{Morales} \& {Hewitt}}{{Morales} \&
  {Hewitt}}{2004}]{Morales:2004}
{Morales} M.~F.,  {Hewitt} J.,  2004, \mn@doi [\apj] {10.1086/424437}, \href
  {http://adsabs.harvard.edu/abs/2004ApJ...615....7M} {615, 7}

\bibitem[\protect\citeauthoryear{{Morales} \& {Wyithe}}{{Morales} \&
  {Wyithe}}{2010}]{Morales:2010}
{Morales} M.~F.,  {Wyithe} J.~S.~B.,  2010, \mn@doi [\araa]
  {10.1146/annurev-astro-081309-130936}, \href
  {http://adsabs.harvard.edu/abs/2010ARA%26A..48..127M} {48, 127}

\bibitem[\protect\citeauthoryear{{Morales}, {Hazelton}, {Sullivan}  \&
  {Beardsley}}{{Morales} et~al.}{2012}]{Morales:2012}
{Morales} M.~F.,  {Hazelton} B.,  {Sullivan} I.,   {Beardsley} A.,  2012,
  \mn@doi [\apj] {10.1088/0004-637X/752/2/137}, \href
  {http://adsabs.harvard.edu/abs/2012ApJ...752..137M} {752, 137}

\bibitem[\protect\citeauthoryear{{Okamoto}, {Gao}  \& {Theuns}}{{Okamoto}
  et~al.}{2008}]{Okamoto:2008}
{Okamoto} T.,  {Gao} L.,   {Theuns} T.,  2008, \mn@doi [\mnras]
  {10.1111/j.1365-2966.2008.13830.x}, \href
  {http://adsabs.harvard.edu/abs/2008MNRAS.390..920O} {390, 920}

\bibitem[\protect\citeauthoryear{{Paciga} et~al.,}{{Paciga}
  et~al.}{2013}]{Pagica:2013}
{Paciga} G.,  et~al., 2013, \mn@doi [\mnras] {10.1093/mnras/stt753}, \href
  {http://adsabs.harvard.edu/abs/2013MNRAS.433..639P} {433, 639}

\bibitem[\protect\citeauthoryear{{Pacucci}, {Mesinger}, {Mineo}  \&
  {Ferrara}}{{Pacucci} et~al.}{2014}]{Pacucci:2014}
{Pacucci} F.,  {Mesinger} A.,  {Mineo} S.,   {Ferrara} A.,  2014, \mn@doi
  [\mnras] {10.1093/mnras/stu1240}, \href
  {http://adsabs.harvard.edu/abs/2014MNRAS.443..678P} {443, 678}

\bibitem[\protect\citeauthoryear{{Parsons} et~al.,}{{Parsons}
  et~al.}{2010}]{Parsons:2010}
{Parsons} A.~R.,  et~al., 2010, \mn@doi [\aj] {10.1088/0004-6256/139/4/1468},
  \href {http://adsabs.harvard.edu/abs/2010AJ....139.1468P} {139, 1468}

\bibitem[\protect\citeauthoryear{{Parsons}, {Pober}, {Aguirre}, {Carilli},
  {Jacobs}  \& {Moore}}{{Parsons} et~al.}{2012}]{Parsons:2012}
{Parsons} A.~R.,  {Pober} J.~C.,  {Aguirre} J.~E.,  {Carilli} C.~L.,  {Jacobs}
  D.~C.,   {Moore} D.~F.,  2012, \mn@doi [\apj] {10.1088/0004-637X/756/2/165},
  \href {http://adsabs.harvard.edu/abs/2012ApJ...756..165P} {756, 165}

\bibitem[\protect\citeauthoryear{{Parsons} et~al.,}{{Parsons}
  et~al.}{2014}]{Parsons:2014}
{Parsons} A.~R.,  et~al., 2014, \mn@doi [\apj] {10.1088/0004-637X/788/2/106},
  \href {http://adsabs.harvard.edu/abs/2014ApJ...788..106P} {788, 106}

\bibitem[\protect\citeauthoryear{{Planck Collaboration} et~al.,}{{Planck
  Collaboration} et~al.}{2015}]{Ade:2015}
{Planck Collaboration} et~al., 2015, preprint, \href
  {http://adsabs.harvard.edu/abs/2015arXiv150201589P} {} (\mn@eprint {arXiv}
  {1502.01589})

\bibitem[\protect\citeauthoryear{{Pober}}{{Pober}}{2015}]{Pober:2015b}
{Pober} J.,  2015, {Updated 21\,cm Experiment Sensitivities},
  \url{http://reionization.org/wp-content/uploads/2015/05/HERA4_sensecalc.pdf}

\bibitem[\protect\citeauthoryear{{Pober} et~al.,}{{Pober}
  et~al.}{2013a}]{Pober:2013b}
{Pober} J.~C.,  et~al., 2013a, \mn@doi [\aj] {10.1088/0004-6256/145/3/65},
  \href {http://adsabs.harvard.edu/abs/2013AJ....145...65P} {145, 65}

\bibitem[\protect\citeauthoryear{{Pober} et~al.,}{{Pober}
  et~al.}{2013b}]{Pober:2013a}
{Pober} J.~C.,  et~al., 2013b, \mn@doi [\apjl] {10.1088/2041-8205/768/2/L36},
  \href {http://adsabs.harvard.edu/abs/2013ApJ...768L..36P} {768, L36}

\bibitem[\protect\citeauthoryear{{Pober} et~al.,}{{Pober}
  et~al.}{2014}]{Pober:2014}
{Pober} J.~C.,  et~al., 2014, \mn@doi [\apj] {10.1088/0004-637X/782/2/66},
  \href {http://adsabs.harvard.edu/abs/2014ApJ...782...66P} {782, 66}

\bibitem[\protect\citeauthoryear{{Pober} et~al.,}{{Pober}
  et~al.}{2015}]{Pober:2015a}
{Pober} J.~C.,  et~al., 2015, preprint, \href
  {http://adsabs.harvard.edu/abs/2015arXiv150300045P} {} (\mn@eprint {arXiv}
  {1503.00045})

\bibitem[\protect\citeauthoryear{{Pritchard} \& {Furlanetto}}{{Pritchard} \&
  {Furlanetto}}{2007}]{Pritchard:2007}
{Pritchard} J.~R.,  {Furlanetto} S.~R.,  2007, \mn@doi [\mnras]
  {10.1111/j.1365-2966.2007.11519.x}, \href
  {http://adsabs.harvard.edu/abs/2007MNRAS.376.1680P} {376, 1680}

\bibitem[\protect\citeauthoryear{{Pritchard} \& {Loeb}}{{Pritchard} \&
  {Loeb}}{2012}]{Pritchard:2012}
{Pritchard} J.~R.,  {Loeb} A.,  2012, \mn@doi [Reports on Progress in Physics]
  {10.1088/0034-4885/75/8/086901}, \href
  {http://adsabs.harvard.edu/abs/2012RPPh...75h6901P} {75, 086901}

\bibitem[\protect\citeauthoryear{{Santos}, {Amblard}, {Pritchard}, {Trac},
  {Cen}  \& {Cooray}}{{Santos} et~al.}{2008}]{Santos:2008}
{Santos} M.~G.,  {Amblard} A.,  {Pritchard} J.,  {Trac} H.,  {Cen} R.,
  {Cooray} A.,  2008, \mn@doi [\apj] {10.1086/592487}, \href
  {http://adsabs.harvard.edu/abs/2008ApJ...689....1S} {689, 1}

\bibitem[\protect\citeauthoryear{{Sobacchi} \& {Mesinger}}{{Sobacchi} \&
  {Mesinger}}{2014}]{Sobacchi:2014}
{Sobacchi} E.,  {Mesinger} A.,  2014, \mn@doi [\mnras] {10.1093/mnras/stu377},
  \href {http://adsabs.harvard.edu/abs/2014MNRAS.440.1662S} {440, 1662}

\bibitem[\protect\citeauthoryear{{Sokolowski} et~al.,}{{Sokolowski}
  et~al.}{2015}]{Sokolowski:2015}
{Sokolowski} M.,  et~al., 2015, \mn@doi [\pasa] {10.1017/pasa.2015.3}, \href
  {http://adsabs.harvard.edu/abs/2015PASA...32....4S} {32, 4}

\bibitem[\protect\citeauthoryear{{Songaila} \& {Cowie}}{{Songaila} \&
  {Cowie}}{2010}]{Songaila:2010}
{Songaila} A.,  {Cowie} L.~L.,  2010, \mn@doi [\apj]
  {10.1088/0004-637X/721/2/1448}, \href
  {http://adsabs.harvard.edu/abs/2010ApJ...721.1448S} {721, 1448}

\bibitem[\protect\citeauthoryear{{Springel} \& {Hernquist}}{{Springel} \&
  {Hernquist}}{2003}]{Springel:2003}
{Springel} V.,  {Hernquist} L.,  2003, \mn@doi [\mnras]
  {10.1046/j.1365-8711.2003.06207.x}, \href
  {http://adsabs.harvard.edu/abs/2003MNRAS.339..312S} {339, 312}

\bibitem[\protect\citeauthoryear{{Tegmark}, {Silk}, {Rees}, {Blanchard}, {Abel}
   \& {Palla}}{{Tegmark} et~al.}{1997a}]{Tegmark:1997c}
{Tegmark} M.,  {Silk} J.,  {Rees} M.~J.,  {Blanchard} A.,  {Abel} T.,   {Palla}
  F.,  1997a, \mn@doi [\apj] {10.1086/303434}, \href
  {http://adsabs.harvard.edu/abs/1997ApJ...474....1T} {474, 1}

\bibitem[\protect\citeauthoryear{{Tegmark}, {Taylor}  \& {Heavens}}{{Tegmark}
  et~al.}{1997b}]{Tegmark:1997d}
{Tegmark} M.,  {Taylor} A.~N.,   {Heavens} A.~F.,  1997b, \apj, \href
  {http://adsabs.harvard.edu/abs/1997ApJ...480...22T} {480, 22}

\bibitem[\protect\citeauthoryear{{Thyagarajan} et~al.,}{{Thyagarajan}
  et~al.}{2013}]{Thyagarajan:2013}
{Thyagarajan} N.,  et~al., 2013, \mn@doi [\apj] {10.1088/0004-637X/776/1/6},
  \href {http://adsabs.harvard.edu/abs/2013ApJ...776....6T} {776, 6}

\bibitem[\protect\citeauthoryear{{Thyagarajan} et~al.,}{{Thyagarajan}
  et~al.}{2015a}]{Thyagarajan:2015a}
{Thyagarajan} N.,  et~al., 2015a, \mn@doi [\apj] {10.1088/0004-637X/804/1/14},
  \href {http://adsabs.harvard.edu/abs/2015ApJ...804...14T} {804, 14}

\bibitem[\protect\citeauthoryear{{Thyagarajan} et~al.,}{{Thyagarajan}
  et~al.}{2015b}]{Thyagarajan:2015b}
{Thyagarajan} N.,  et~al., 2015b, \mn@doi [\apjl]
  {10.1088/2041-8205/807/2/L28}, \href
  {http://adsabs.harvard.edu/abs/2015ApJ...807L..28T} {807, L28}

\bibitem[\protect\citeauthoryear{{Tingay} et~al.,}{{Tingay}
  et~al.}{2013}]{Tingay:2013a}
{Tingay} S.~J.,  et~al., 2013, \mn@doi [\pasa] {10.1017/pasa.2012.007}, \href
  {http://adsabs.harvard.edu/abs/2013PASA...30....7T} {30, 7}

\bibitem[\protect\citeauthoryear{{Trott}, {Wayth}  \& {Tingay}}{{Trott}
  et~al.}{2012}]{Trott:2012}
{Trott} C.~M.,  {Wayth} R.~B.,   {Tingay} S.~J.,  2012, \mn@doi [\apj]
  {10.1088/0004-637X/757/1/101}, \href
  {http://adsabs.harvard.edu/abs/2012ApJ...757..101T} {757, 101}

\bibitem[\protect\citeauthoryear{{Vedantham}, {Udaya Shankar}  \&
  {Subrahmanyan}}{{Vedantham} et~al.}{2012}]{Vedantham:2012}
{Vedantham} H.,  {Udaya Shankar} N.,   {Subrahmanyan} R.,  2012, \mn@doi [\apj]
  {10.1088/0004-637X/745/2/176}, \href
  {http://adsabs.harvard.edu/abs/2012ApJ...745..176V} {745, 176}

\bibitem[\protect\citeauthoryear{{Vogeley} \& {Szalay}}{{Vogeley} \&
  {Szalay}}{1996}]{Vogeley:1996}
{Vogeley} M.~S.,  {Szalay} A.~S.,  1996, \mn@doi [\apj] {10.1086/177399}, \href
  {http://adsabs.harvard.edu/abs/1996ApJ...465...34V} {465, 34}

\bibitem[\protect\citeauthoryear{{Voytek}, {Natarajan}, {J{\'a}uregui
  Garc{\'{\i}}a}, {Peterson}  \& {L{\'o}pez-Cruz}}{{Voytek}
  et~al.}{2014}]{Voytek:2014}
{Voytek} T.~C.,  {Natarajan} A.,  {J{\'a}uregui Garc{\'{\i}}a} J.~M.,
  {Peterson} J.~B.,   {L{\'o}pez-Cruz} O.,  2014, \mn@doi [\apjl]
  {10.1088/2041-8205/782/1/L9}, \href
  {http://adsabs.harvard.edu/abs/2014ApJ...782L...9V} {782, L9}

\bibitem[\protect\citeauthoryear{{Xue} et~al.,}{{Xue} et~al.}{2011}]{Xue:2011}
{Xue} Y.~Q.,  et~al., 2011, \mn@doi [\apjs] {10.1088/0067-0049/195/1/10}, \href
  {http://adsabs.harvard.edu/abs/2011ApJS..195...10X} {195, 10}

\bibitem[\protect\citeauthoryear{{van Haarlem} et~al.,}{{van Haarlem}
  et~al.}{2013}]{VanHaarlem:2013}
{van Haarlem} M.~P.,  et~al., 2013, \mn@doi [\aap]
  {10.1051/0004-6361/201220873}, \href
  {http://adsabs.harvard.edu/abs/2013A%26A...556A...2V} {556, A2}

\makeatother
\end{thebibliography}

\appendix

\end{document}